\documentclass{raa}

\usepackage{graphicx,times}             

\usepackage{natbib} 
\usepackage{amsmath,amssymb}

\newcommand{\hMsol}{{\>h^{-1}\rm M}_\odot}
\newcommand{\hMpc}{{\>h^{-1}\rm Mpc}}
\newcommand{\hkpc}{{\>h^{-1}\rm kpc}}
\newcommand{\kms}{\>{\rm km}\,{\rm s}^{-1}}

\newcommand{\lra}[1]{{\langle #1\rangle}}
\renewcommand{\vec}[1]{{\mathbf #1}}

\begin{document}

   \title{Alignment between galaxies and large-scale structure
}

   \volnopage{Vol.0 (200x) No.0, 000--000}      
   \setcounter{page}{1}          


   \author{A.   Faltenbacher\inst{1,2}
     \and 
     Cheng Li\inst{1,2}  
     \and 
     Simon D.~M. White\inst{1}
     \and 
     Y.~P. Jing\inst{2}
     \and Shude Mao\inst{3}
     \and 
     Jie Wang\inst{4}
   }

\institute{
  Max-Planck-Institute            for           Astrophysics,
  Karl-Schwarzschild-Str.    1,   D-85741   Garching,   Germany   {\it
    afaltenbacher@mpa-garching.mpg.de}\\ 
  \and
  MPA/SHAO  Joint  Center  for  Astrophysical  Cosmology  at  Shanghai
  Astronomical Observatory, Nandan Road 80, Shanghai 200030, China\\
  \and
  Jodrell  Bank Centre  for  Astrophysics, Alan  Turing Building,  The
  University of Manchester, Manchester M13 9PL\\ 
  \and
  Department of Physics, Institute of Computational Cosmology, University of Durham,
  Science Laboratories, South Road, Durham DH1 3LE
}

\date{Received~~2008 month day; accepted~~2008~~month day}

\abstract{  Based on  the  Sloan  Digital Sky  Survey  DR6 (SDSS)  and
  Millennium  Simulation  (MS) we  investigate  the alignment  between
  galaxies and large-scale structure.  For this purpose we develop two
  new  statistical  tools,   namely  the  {\it  alignment  correlation
    function} and the  {\it $\cos(2\theta)$-statistic}.  The former is
  a two-dimensional extension of the traditional two-point correlation
  function and  the latter is  related to the  ellipticity correlation
  function used for cosmic shear  measurements.  Both are based on the
  cross correlation between a sample of galaxies with orientations and
  a reference  sample which  represents the large-scale  structure. We
  apply the new statistics to  the SDSS galaxy catalog.  The alignment
  correlation function reveals  an overabundance of reference galaxies
  along the  major axes of red, luminous  ($L\gtrsim L_\ast$) galaxies
  out  to projected  separations of  $60\hMpc$.  The  signal increases
  with central galaxy luminosity.  No alignment signal is detected for
  blue  galaxies.  The  $\cos(2\theta)$-statistic yields  very similar
  results. Starting  from a MS semi-analytic galaxy  catalog we assign
  an orientation  to each red,  luminous and central galaxy,  based on
  that of  the central region of  the host halo (with  size similar to
  that  of  the  stellar  galaxy).   As  an  alternative  we  use  the
  orientation  of the  host  halo  itself. We  find  a mean  projected
  misalignment  between  a  halo  and  its  central  region  of  $\sim
  25^\circ$.   The  misalignment  decreases slightly  with  increasing
  luminosity  of  the  central  galaxy.  Using  the  orientations  and
  luminosities of  the semi-analytic galaxies we  repeat our alignment
  analysis on mock surveys of the MS.  Agreement with the SDSS results
  is good if the central orientations are used.  Predictions using the
  halo  orientations  as   proxies  for  central  galaxy  orientations
  overestimate  the observed  alignment by  more than  a factor  of 2.
  Finally,  the  large  volume of  the  MS  allows  us to  generate  a
  two-dimensional  map  of the  alignment  correlation function  which
  shows the reference galaxy  distribution to be flattened parallel to
  the  orientations  of red  luminous  galaxies  with  axis ratios  of
  $\sim0.5$  and   $\sim0.75$  for  halo   and  central  orientations,
  respectively.  These  ratios are almost independent of  scale out to
  $60\hMpc$.  }

   \authorrunning{A. Faltenbacher et al.}            
   \titlerunning{Alignment between galaxies and large-scale structure}  

   \maketitle

%
%
\section{Introduction}
\label{sec:intro}
Recent  large redshift surveys,  like the  2dF Galaxy  Redshift Survey
\citep[2dFGRS,][]{Colless-01}  and   the  Sloan  Digital   Sky  Survey
\citep[SDSS,][]{York-00}, allow  the cosmic large-scale  density field
to be  traced with  unprecedented accuracy.  Different  structures can
reliably be classified as groups, filaments, walls and voids and match
well with  the patterns seen in N-body  simulations.  These structures
induce  large-scale  tidal  fields  which in  turn  cause  large-scale
correlations  in  the  orientations   of  massive  dark  matter  halos
\citep[cf.][]{Bond-Kofman-Pogosyan-96,                      Colberg-99,
  Altay-Colberg-Croft-06} .  However,  the orientations of dark matter
halos are  difficult to observe.  One needs  either X-ray observations
for a sufficient number of groups and clusters of galaxies or reliable
galaxy-group catalogs  derived from the redshift  surveys.  The former
are  expensive and  the latter  are prone  to a  number  of systematic
errors.  Despite these difficulties both approaches have been pursued.
In most cases alignment signals  out to, at least, $20\hMpc$ have been
found          \citep{Binggeli-82,         Ulmer-McMillan-Kowalski-89,
  West-89a,Plionis-94,                       Chambers-Melott-Miller-00,
  Hashimoto-Henry-Bohringer-07}.  Instead of measuring the large-scale
alignment based on groups and  clusters of galaxies we suggest here to
use  directly the orientations  of luminous  galaxies and  we quantify
their alignment relative to large-scale structure.

Galaxies are not oriented at  random.  Rather, they have been found to
show various forms of  spatial alignment: between neighboring clusters
of  galaxies  \cite[][]{Binggeli-82,  West-89b,  Plionis-94},  between
brightest   cluster  galaxies   (BCGs)  and   their   parent  clusters
\cite[][]{Carter-Metcalfe-80,         Binggeli-82,         Struble-90,
  Hashimoto-Henry-Boehringer-08}, between the orientation of satellite
galaxies  and  the  orientation  of  the  cluster  \cite[][]{Dekel-85,
  Plionis-03}, and  between the orientation of  satellite galaxies and
the orientation of the BCG \cite[][]{Struble-90}.

Observationally,  these  alignments   are  quantified  either  by  the
differential, $P(\theta)$, or cumulative, $P(\theta\leq\theta_{\rm max})$,
probability distribution of the  alignment angle $\theta$ which is the
angle between the major axis of a galaxy and the line connecting it to
a neighboring  galaxy.  Also the  mean values of  those distributions,
$\langle\theta\rangle(r_p)$,  have  been  studied  as  a  function  of
projected separation  $r_p$.  With  recent large redshift  surveys, in
particular the SDSS, it has become possible to determine the alignment
using large  and homogeneous samples.  Studies based  on these surveys
have  focused mainly  on the  alignment of  galaxies in  groups.  They
revealed that satellite  galaxies are preferentially distributed along
the  major   axis  of  the   central  galaxies  \cite[][]{Brainerd-05,
  Yang-06b, Azzaro-07,  Faltenbacher-07}, and satellite  galaxies tend
to   be   preferentially    oriented   toward   the   central   galaxy
\cite[][]{Pereira-Kuhn-05,  Agustsson-Brainerd-06b,  Faltenbacher-07}.
\cite{Donoso-O'Mill-Lambas-06} analyzed a high-redshift sample ($0.4 <
z < 0.5$) of luminous red galaxies (LRGs) extracted from the SDSS, and
found  a clear  signal of  alignment between  LRG major  axes  and the
distribution of  galaxies within $1.5 h^{-1}$Mpc,  indicating that the
alignment effects observed in  the local Universe were already present
at $z\sim 0.5$.

In this paper we propose  two new statistics to quantify the alignment
between galaxies and the large-scale structure.  The first one we call
the  {\em  alignment  two-point  correlation  function}  which  is  an
extension     of     the     traditional     two-point     correlation
function. Basically, the correlation is measured as a function of pair
separation  and alignment  angle. The  second  measure we  call  {\em
  $cos(2\theta)$-statistic} and  involves determining the  mean cosine
of twice the  alignment angle for correlated pairs  of given projected
separation. We measure  these statistics for SDSS galaxies  as well as
for  semi-analytic galaxies  within the  Millennium  simulation (MS)
\citep{Springel-05b,  DeLucia-Blaizot-07}, where  the  orientations of
the semi-analytic galaxies are  inferred from the orientation of their
parent  dark matter  halos.   The application  of  the new  statistics
reveals an  alignment between  red galaxies and  large-scale structure
out to $60\hMpc$.

The paper  is organized as  follows. In \S~\ref{sec:met}  we introduce
the  two  statistics  used  here  to quantify  the  alignment  between
galaxies and large-scale structure.  The statistics are applied to the
SDSS galaxy  catalog in \S~\ref{sec:sdss} and compared  to the results
derived  from   the  semi-analytic  galaxy   catalog  of  the   MS  in
\S~\ref{sec:mille}.  Finally, \S~\ref{sec:sum} gives a short summary.
\section{Methodology}
\label{sec:met}
For the subsequent analysis  we introduce two new statistical measures
to  quantify  the correlations  between  galaxy  orientations and  the
cosmic density field.   The first quantity we call  the {\it alignment
  (two-point)   correlation   function},  $w_p(\theta_p,r_p)$,   where
$\theta_p$ is  the angle between  the major axis  of a galaxy  and the
connecting line to  another one and $r_p$ is  the projected separation
between  the  two  galaxies.    This  quantity  is  a  two-dimensional
extension of the traditional two-point projected correlation function.
\cite{Paz-Stasyszyn-Padilla-08}  have  used   a  related  technique to
analyze large-scale angular  momentum alignments.  The second quantity
we refer to as  the {\em $cos(2\theta)$-statistic}.  This statistic is
closely related  to similar quantities  used in the context  of cosmic
shear surveys.   We compute both  statistics only in  projected space,
i.e.  orientations  and pair separations  are two-dimensional vectors.
We  keep this restriction  also in  the second  part of  this analysis
where we compare observational and numerical results.
\subsection{Alignment correlation function}
\label{sec:ali2p}
The  two-point correlation  function  (2PCF) has  long  served as  the
primary way  of quantifying the  clustering properties of  galaxies in
redshift  surveys  \cite[e.g.][]{Peebles-80}.   It  is  defined  as  a
function of pair separation by
\begin{equation}
  dP_{12} = \bar{n}^2[1+\xi(\vec{r})]dV_1dV_2,
\end{equation}
where $\xi(\vec{r})$ is the 2PCF, $\bar{n}$ the mean number density of
galaxies, and $dV_1$ and  $dV_2$ are two infinitesimal volume elements
centered    at   $\vec{x}_1$    and   $\vec{x}_2$    with   separation
$\vec{r}=\vec{x}_2-\vec{x}_1$.  To be  consistent with homogeneity and
isotropy $\xi$ has been written as a function of the separation alone,
that  is, $\xi(r)$.   If $\xi(r)\ne0$,  then galaxies  are said  to be
clustered.  As  the fundamental second-order statistic  of the density
field,  $\xi$ is  simple to  compute and  provides a  full statistical
description for a Gaussian random field \cite[][]{Bardeen-86}.  It can
also  be easily compared  with the  predictions of  theoretical models
\cite[e.g.][]{Aarseth-Turner-Gott-79,Davis-85}.  The  amplitude of the
correlation function on scales larger than a few Mpc provides a direct
measure of  the mass of the  dark matter halos that  host the galaxies
through          the          halo         mass-bias          relation
\cite[e.g.][]{Mo-White-96,Jing-Mo-Boerner-98}.

In galaxy redshift surveys, the 2PCF is measured in redshift space and
expressed  either  as a  function  of  redshift-space separation  $s$,
giving  rise to a  2PCF of  $\xi(s)$, or  as functions  of separations
perpendicular  ($r_p$) and  parallel  ($\pi$) to  the  line of  sight,
giving rise  to $\xi(r_p,\pi)$ with $s^2=\pi^2+r_p^2$.  In many cases,
the projected  2PCF, $w_p(r_p)$,  is the more  useful quantity,  as it
does not suffer from  redshift-space distortions, and is thus directly
related  to the  real-space  correlation function  $\xi(r)$.  One  can
distinguish between two kinds of two-point correlation functions: {\em
  two-point auto-correlation  functions} for  which both members  of a
pair come  from the same sample, and  {\em two-point cross-correlation
  functions} (2PCCFs) for which the two members of a pair are from two
different samples.  In this paper we focus on the latter.  In addition
our analysis will be pursued in projected space, i.e.  we focus on the
projected   2PCCF  $w_p(r_p)$  where   $r_p$  denotes   the  projected
separation of a galaxy pair.

Given  a sample  of galaxies  in question  (Sample $Q$),  a  sample of
reference  galaxies   (Sample  $R$),  and  a   random  sample  (Sample
$\mathcal{R}$) that has the same selection function (i.e. distribution
of redshifts  and positions on the  sky) as the  reference sample,
$\xi(r_p,\pi)$ between $Q$ and $R$ can be estimated by
\begin{equation}\label{eqn:xipv}
  \xi(r_p,\pi)=
  \frac{N_{\mathcal{R}}}{N_{R}}\frac{QR(r_p,\pi)}{Q\mathcal{R} (r_p,\pi)} - 1,
\end{equation}
where  $N_{R}$  and  $N_{\mathcal{R}}$  are  the  number  of  galaxies
contained    in    the   reference    and    random   samples,    with
$N_{\mathcal{R}}/N_{R}=10$  throughout this paper.   $QR(r_p,\pi)$ and
$Q\mathcal{R}(r_p,\pi)$  are the  counts  of cross  pairs between  the
indicated  samples for  a given  separation perpendicular,  $r_p$, and
parallel,  $\pi$,  to  the  line-of-sight.  With  the  measurement  of
$\xi(r_p,\pi)$   in  hand   one  can   then  estimate   the  projected
cross-correlation  function $w_p(r_p)$  by  integrating $\xi(r_p,\pi)$
along the $\pi$ direction:
\begin{equation}\label{eqn:wrp}
  w_p(r_p)    =
  \int_{-\pi_{\rm max}}^{+\pi_{\rm max}}\xi(r_p,\pi)d\pi    =
  \sum_{i}\xi(r_p,\pi_i)\Delta\pi_i,
\end{equation}
where $\pi_{\rm max}$ has  to be sufficiently  large to  minimize the
probability   of   erroneously   excluding   correlated   pairs   with
line-of-sight separations larger than $\pi_{\rm max}$.

Now we extend the definition of  the 2PCCFs, so that they will be able
to  quantify the  spatial alignment  of  galaxies.  For  each pair  of
galaxies with  one member  from Sample $Q$  (the main galaxy)  and the
other from Sample $R$  (the reference galaxy), we consider $\theta_p$,
the  angle between  the major  axis of  the main  galaxy and  the line
connecting the two  galaxies projected onto the sky.   We include this
angle  as a  second property  of  the pair,  in addition  to the  pair
separation.   In this  case, the  correlation function  is not  only a
function of  the projected separations,  but also of  $\theta_p$.  The
estimator of  Eq.  (\ref{eqn:xipv}) is easily modified  to account for
the dependence on $\theta_p$:
\begin{equation}\label{eqn:xipv}
  \xi(\theta_p,r_p,\pi)=
  \frac{N_{\mathcal{R}}}{N_{R}}\frac{QR(\theta_p,r_p,\pi)}{Q\mathcal{R} 
    (\theta_p,r_p,\pi)} - 1,
\end{equation}
$QR(\theta_p,r_p,\pi)$  and  $Q\mathcal{R}(\theta_p,r_p,\pi)$ are  the
counts  of  cross  pairs  between  the  indicated  samples  for  given
$\theta_p$,  $r_p$ and  $\pi$. The  projected correlation  function is
found by integration along the line-of-sight.
\begin{equation}
  \label{eqn:cwrp}
  w_p(\theta_p,r_p) =
\int_{-\pi_{\rm max}}^{+\pi_{\rm max}}\xi(\theta_p,r_p,\pi)d\pi =
\sum_{i}\xi(\theta_p,r_p,\pi_i)\Delta\pi_i
\end{equation}
The traditional  correlation function is  just the average of  the new
correlation function over the full range of $\theta_p$ values.  Taking
symmetries into account the value of the angle ranges from zero (along
the major axis of the main galaxy) to 90 degrees (perpendicular to the
major axis).  Thus, higher amplitudes of the new correlation functions
at small values of $\theta_p$ indicate that the reference galaxies are
more likely to  be aligned along the major axis  of the main galaxies.
In  contrast, higher  amplitudes at  larger angles  indicate  that the
reference galaxies are more likely  to be located along the minor axis
of the main galaxies.  This  new statistic can be used for quantifying
the alignment  of galaxies and  we refer to  it as the  {\em alignment
  correlation function}.

On small scales ($\la$ Mpc), this statistic can be used to confirm the
alignment  between  central  and  satellite  galaxies  in  groups  and
clusters,  if   Sample  $Q$   consists  purely  of   central  galaxies
\citep[cf.][]{Carter-Metcalfe-80,Binggeli-82,    Struble-90}.     More
interestingly,  the new statistic  allows us  to extend  the alignment
study to  very large scales without worrying  about selection effects,
which  are taken  into account  by comparison  with the  random sample
($\mathcal{R}$).   On large  scales,  this statistic  can  be used  to
quantify the alignment  of the main galaxies (which may  or may not be
central  galaxies) with respect  to the  large-scale structure  of the
Universe as  probed by the  large-scale distribution of  the reference
galaxies.

As the final remark of this  section we would like to mention that one
can  easily  derive  equivalent  expressions for  the  alignment  {\em
  auto-}correlation function --  additionally it is straightforward to
extend the formalism to three dimensional problems.
\subsection{The $\cos(2\theta)$-statistic}
\label{sec:cos2theta}
The $\cos(2\theta)$-statistic  gives the average  of $\cos(2\theta_p)$
for all correlated pairs at  a given projected separation.  It will be
referred as  $\lra{\cos(2\theta_p)_{cor}}(r_p)$ where the  index $cor$
emphasizes that the average is based on correlated pairs only.  Again,
$\theta_p$ indicates the angle between the major axis of a main galaxy
and the  line connecting it  with a reference galaxy.  More precisely,
using  the  alignment  correlation function,  $w_p(\theta_p,r_p)$,  we
define 
\begin{equation}
\label{eqn:cos2theta}
\lra{\cos(2\theta_p)_{cor}}(r_p)
= {\int_0^{\pi/2}\cos(2\theta_p)w_p(\theta_p,r_p)d\theta_p
\over
\int_0^{\pi/2}w_p(\theta_p,r_p)d\theta_p}\ . 
\end{equation}
This statistic is constrained to  values between $-1$ and $1$.  Values
above and  below 0 indicate  a preference for small  ($<45^\circ$) and
large  ($>45^\circ$)  angles, respectively.   Values  close  to 0  are
expected  for isotropy.  An estimator  for  Eq.~\ref{eqn:cos2theta} is
given by
\begin{equation} 
\label{eqn:co2est}
\lra{\cos(2\theta_p)_{cor}}(r_p)=
  {\sum_i
  \left({QR_\theta(r_p,\pi_i)\over
    Q\mathcal{R}(r_p,\pi_i)}\right) \Delta\pi_i 
  \over
  \sum_i \left({QR(r_p,\pi_i)\over Q\mathcal{R}(r_p,\pi_i)}-1\right)\Delta\pi_i
  }\ ,
 \end{equation}
where $QR_\theta(r_p,\pi_i)$  symbolizes the sum  of $\cos(2\theta_p)$
for all cross  pairs between main and reference  samples ($Q$ and $R$)
within  the given  separation  bins, $r_p$  and  $\pi_i$.  As  before,
$QR(r_p,\pi_i)$  and $Q\mathcal{R}(r_p,\pi_i)$  denote  the number  of
cross  pairs between  the  indicated samples  for  $r_p$ and  $\pi_i$.

Related statistics  are used  in weak lensing  studies where  they are
referred    to     as    ellipticity    correlations    \citep[e.g.][]
{Miralda-Escude-91,Croft-Metzler-00,Heavens-Refregier-Heymans-00}.  In
particular we  want to point out  the similarity to  the the intrinsic
shear-density         correlation        function        $w_{g+}(r_p)$
\citep[][]{Mandelbaum-06a,  Hirata-07} which measures  the correlation
between galaxy orientations  and the large-scale density distribution.
Leaving the ellipticity  weighting and `responsivity' correction aside
(cf.  Eqs.~8 and 9 in ~\citealt{Hirata-07}) the following relation holds.
\begin{equation}
  \tilde{w}_{g+} = w_p(r_p) \lra{\cos(2\theta_p)}_{cor}(r_p)
\end{equation}
where $\tilde{w}_{g+}$  indicates the unweighted  version of $w_{g+}$.
We do not weight by ellipticity since we here are solely interested in
the spatial alignment between galaxies and the large-scale structure.
\section{Alignment of SDSS galaxies}
\label{sec:sdss}
In  this section  we  apply the  two  new statistics  to the  publicly
available data  from the Sloan  Digital Sky Survey DR6.   Before doing
that  we describe some  details of  the survey  and the  galaxy sample
construction.  Also  the determination  of the galaxy  orientations is
reviewed.
\subsection{Galaxy sample construction}
\label{sec:sdss.sample}
The  observational data used  in this  paper are  taken from  the SDSS
which has been  designed to obtain photometry of a  quarter of the sky
and spectra of nearly one million objects.  Imaging is obtained in the
{\em u,  g, r, i,  z} bands \cite[][]{Fukugita-96,Smith-02,Ivezic-04}.
The details of the survey  strategy can be found in \cite{York-00} and
an  overview of the  data pipelines  and products  is provided  in the
Early Data  Release paper  \cite[][]{Stoughton-02}.  The SDSS  has had
seven   additional   major   data   releases   \cite[][]{Abazajian-03,
  Abazajian-04,
  Abazajian-05,Adelman-McCarthy-06,Adelman-McCarthy-07,Adelman-McCarthy-08}.

The galaxy samples for this work are constructed from {\tt Sample dr6}
of the  New York University  Value Added Galaxy  Catalogue (NYU-VAGC),
which   is   based   on   the   SDSS  DR6,   publicly   available   at
http://sdss.physics.nyu.edu/vagc/. A  detailed description thereof can
be found in \cite{Blanton-05}.  Our sample consists of 430164 galaxies
that  are identified  as  galaxies  from the  Main  sample (note  that
$r$-band magnitude has been  corrected for foreground extinction), and
are   in  the  ranges   of  $0.01<z<0.4$,   $-23<M_{^{0.1}r}<-17$  and
$14.5<r<17.6$.  Here $M_{^{0.1}r}$  is the $r$-band absolute magnitude
corrected  to its $z  = 0.1$  value using  the $K-$correction  code of
\cite{Blanton-03a}   and    the   luminosity   evolution    model   of
\cite{Blanton-03b}.   We   do  not  consider   galaxies  fainter  than
$M_{^{0.1}r}=-17$, because the volume covered by such faint samples is
very small and the results are  subject to large errors as a result of
cosmic  variance \cite[see  for  example Fig.   6 of][]{Li-06b}.   The
faint apparent magnitude  limit of 17.6 is chosen  to yield an uniform
galaxy sample  that is  complete over the  entire area of  the survey.
This  sample will  serve  as  the reference  sample  (Sample $R$)  for
computing cross-correlation  functions, as  well as the  parent sample
for selecting different subsamples ($Q$).

The large  amount of data  allows us to  split the parent  sample into
various subsamples.  Thus, we split all the galaxies into 4 subsamples
according  to   their  $r$-band  absolute   magnitudes,  ranging  from
$M_{^{0.1}r}=-23$   to  $M_{^{0.1}r}=-17$  with   an  interval   of  1
magnitude.  Further, we classify each galaxy with $M_{^{0.1}r}>-22$ as
either ``red'' or ``blue'' according to its $g-r$ color.  To this end,
we  follow  \cite{Li-06b}  to  fit  the $g-r$  distribution  at  fixed
luminosity  with a bi-Gaussian  profile and  use the  mean of  the two
Gaussian  centers as  the color  cut. In  the highest  luminosity bin,
$-23<M_{^{0.1}r}<-22$,  this color  separation scheme  is problematic,
basically  because the blue  population is  extremely sparse,  thus we
assign all  galaxies within this  magnitude bin to the  red population
\citep[cf.][Fig.~9]{Li-06b}.      These     subsamples     will     be
cross-correlated  with  the full  parent  sample,  i.e. the  reference
sample.

To  compute  the  cross-correlation  function  between  the  main  and
reference samples, $Q$  and $R$, one also needs  to construct a random
sample, $\cal  R$, where galaxies with random  coordinates are subject
to  the same  selection effects  as the  reference sample.   Since the
reference sample is used as a tracer of the large-scale structure only
the galaxy  positions instead of  their orientations are  of interest.
Accordingly,  orientations are  not considered  when  constructing the
random sample. A detailed account of the selection effects accompanies
the NYU-VAGC  release \cite[][]{Blanton-05} which is the  base for the
construction of the random sample used here \citep[cf.][]{Li-06b}.
\subsection{Determination of galaxy orientations}
\label{sec:sdss.ori}
The orientation of  galaxies is quantified by the  position angle (PA)
of the major axis of their $r$-band images, which is determined by the
SDSS    photometric     pipeline    called    {\tt     {\sc    Photo}}
\cite[][]{Lupton-01}. {\tt {\sc  Photo}} provides three quantities for
the  PA of each  galaxy: PA$_{deV}$,  PA$_{exp}$, and  PA$_{iso}$. The
first two come from fitting two models to the two-dimensional image of
the galaxy  in each  band: a  pure de Vaucouleurs  profile and  a pure
exponential  profile, while  the last  one is  given by  measuring the
shape  parameters (centroid,  major and  minor axes,  PA,  and average
radius)  of  the  25  mag  arcsec$^{-2}$  isophote.   Details  on  the
photometric   pipeline   can   be   found  in   \cite{Lupton-01}   and
\cite{Stoughton-02}.   In   this  paper  we  use   the  isophotal  PA,
PA$_{iso}$.   However,  we  found  no  significant  variation  in  the
alignment signals  when adopting  the alternative definitions  for the
PAs.

To quantify the intrinsic scatter  of the alignment signal we randomly
shuffle the  PAs among the  galaxies in the  main sample and  redo the
alignment   analysis.   Since  the   shuffled  orientations   are  not
correlated with  the large-scale  structure one expects  no systematic
alignment  signal.  The  $1\sigma$  variance of  10 randomly  shuffled
samples can then be used to infer the intrinsic scatter.
\subsection{Alignment correlation function}
\label{sec:sdss.ali}
\begin{figure}
  \centering
  \includegraphics[width=0.8\textwidth]{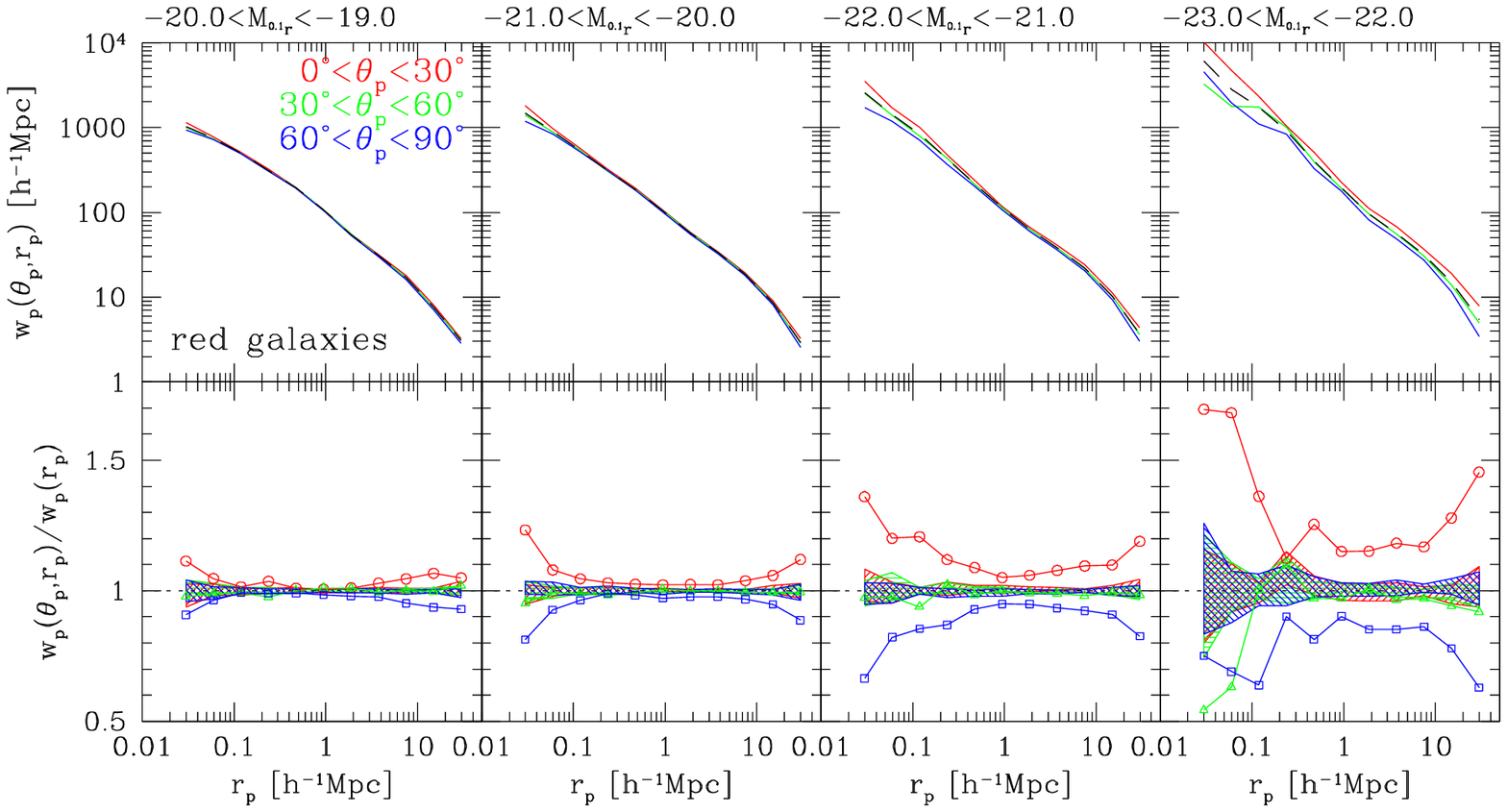}
  \\[0.03\vsize]
  \includegraphics[width=0.8\textwidth]{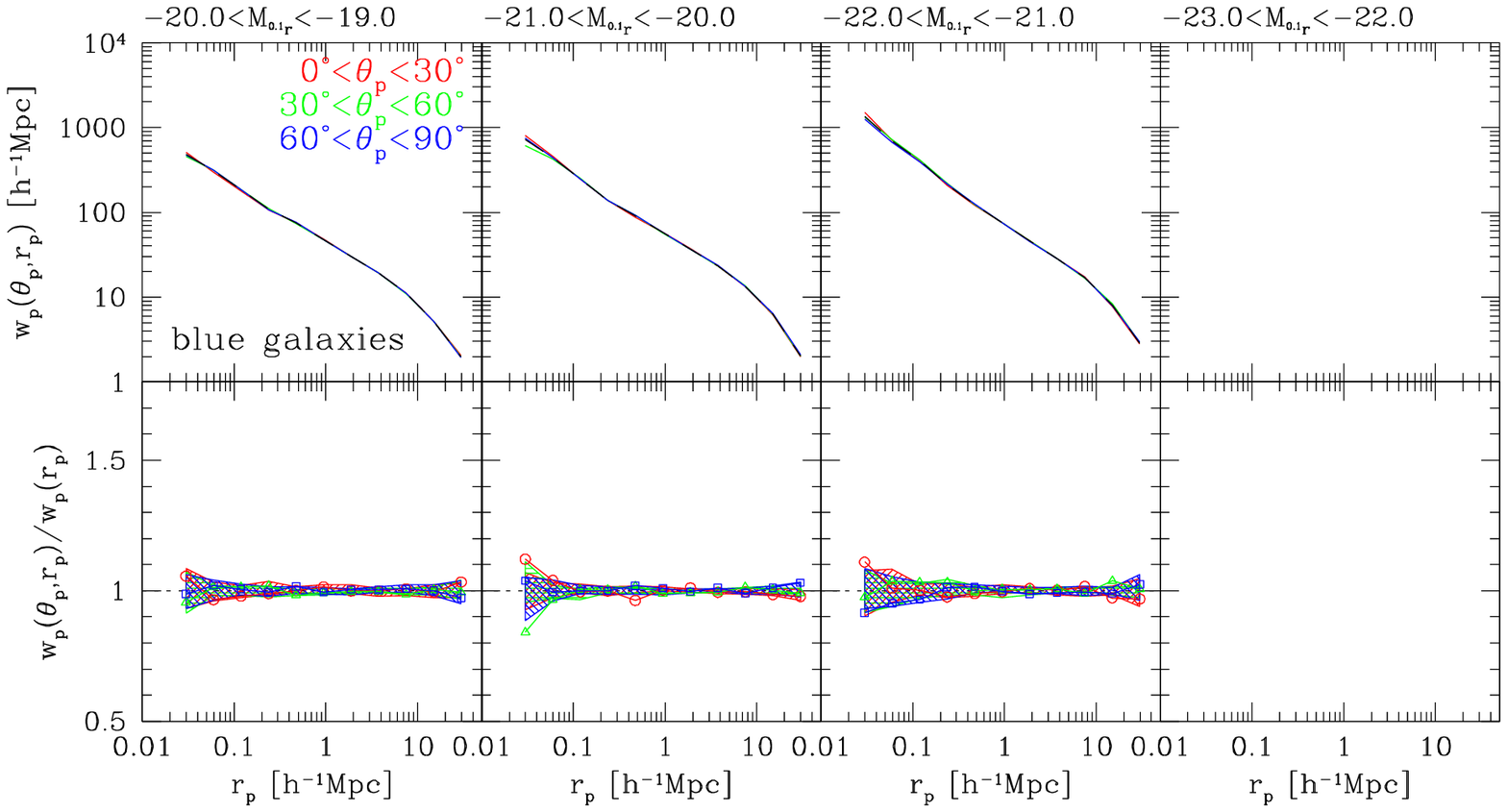}
\caption{\label{fig:wrp} {\it  Upper panels in each  set:} Solid lines
  display   the  projected   SDSS   alignment  correlation   function,
  $w_p(\theta_p,r_p)$,  between our  reference sample  and  red (upper
  set) and  blue (lower set)  main galaxies in different  intervals of
  $r$-band absolute magnitude.  The  colors correspond to three ranges
  in  $\theta_p$  as  indicated.   The  dashed black  line  shows  the
  conditional  correlation function averaged  over angle.   {\it Lower
    panels in each set:} Ratio between the angle dependent correlation
  functions and their mean.  The color code is the same as used in the
  upper panels.   The shaded  regions plotted in  red, green  and blue
  colors indicate the $1-\sigma$ variance based on 10 samples in which
  the position angles  are shuffled at random among  the galaxies.  By
  definition (\S~\ref{sec:sdss.sample}) there  are no blue galaxies in
  the brightest luminosity bin so these panels are kept void.}
\end{figure}

We    now     compute    the    alignment     correlation    function,
$w_p(\theta_p,r_p)$,     based      on     Eq.~\ref{eqn:cwrp}     with
$\pi_{\rm max}=40\hMpc$.  We  probe projected separations  up to $60\hMpc$
within  three  angular  bins.    A  less  coarse  angular  binning  is
prohibited by Poisson noise at large separations.
\subsubsection{Alignment for red galaxies}
\label{sec:sdss.ali.red}
The  first row  of the  upper  panel of  Fig.~\ref{fig:wrp} shows  the
alignment correlation  function for red galaxies  in various magnitude
bins. The red, green and blue  solid lines display the results for the
different  angular bins,  $0^\circ-30^\circ$,  $30^\circ-60^\circ$ and
$60^\circ-90^\circ$,    respectively.      The    low    angle    bin,
$0^\circ-30^\circ$,  quantifies  the  correlation function  along  the
orientation  of  the   main  galaxies.   The  $60^\circ-90^\circ$  bin
contains   information   about    the   reference   galaxy   abundance
perpendicular  to the  orientation  of the  main  galaxies.  The  long
dashed black line represents  the traditional correlation function and
is a weighted average of the three bins.

The  second row  of the  upper panel  of Fig.~\ref{fig:wrp}  shows the
ratio  of  the  alignment  correlation  function  to  the  traditional
one. The color code  is the same as used in the  row above. The shaded
regions indicate the $1 \sigma$  variance between 10 random samples in
which the  orientations are shuffled  at random among the  galaxies. A
signal well  outside the shaded  region is a significant  detection of
alignment or anti-alignment depending on whether the signal lies above
or below one.

The alignment  correlation functions (colored lines)  for red galaxies
in  the lowest  luminosity bin  ($-20<M_{^{0.1}r}<-19$) can  hardly be
distinguished from the  traditional correlation function (dashed black
line).         However,         the        corresponding        ratio,
$w_p(\theta_p,r_p)/w_p(r_p)$,  shows a  weak but  systematic  trend to
anisotropy at scales larger  than $\sim10\hMpc$.  At these separations
the reference galaxies are preferentially located along the major axes
of the main galaxies.  Along the minor axes the reference galaxies are
correspondingly under-abundant.  This  feature becomes more pronounced
in   higher    luminosity   bins.    For    galaxies   brighter   than
$M_{^{0.1}r}=-20$ a significant overabundance of reference galaxies is
visible along the  major axis.  The signal reaches  from $10\hkpc$ out
to $60\hMpc$ which corresponds the entire range probed here.

These  plots  indicate  that  the  orientations of  red  galaxies  are
connected  to the large-scale  structure in  which they  are embedded.
Any realistic galaxy formation model  should be able to reproduce this
interlacing between dimensions spanning $\sim3$ orders of magnitude.
\subsubsection{Alignment for blue galaxies}
\label{sec:sdss.ali.blue}
The  lower  panels  of  Fig.~\ref{fig:wrp} display  the  corresponding
results for blue  galaxies.  Otherwise the analysis is  carried out in
exactly the  same manner as for  the red galaxies.  The  panel for the
highest magnitude bin, $-23<M_{^{0.1}r}<-22$,  is left void because by
definition  there are no  blue galaxies  with these  luminosities (see
\S~\ref{sec:sdss.sample}).  In  the lower  three magnitude bins  we do
not find  any indication  for alignment between  the galaxies  and the
reference galaxy  distribution.  The differences seen  between red and
blue galaxies suggest  that the orientations of red  and blue galaxies
are determined by distinct physical processes.
\subsection{$\cos(2\theta)$-statistic}
\label{sec:sdss.cos}
In this section we compute  the $\cos(2\theta)$ statistic based on the
same  subdivision of  the parent  SDSS  sample by  color and  $r$-band
absolute  magnitude  as  used  above  for  the  alignment  correlation
function.  In a second step we focus on the alignment signal for group
central galaxies.  This is  to facilitate a comparison with simulation
results discussed in the second part of this paper.
\subsubsection{Color and luminosity dependence}
\label{sec:sdss.cos.all}
\begin{figure}
  \includegraphics[width=0.95\textwidth]{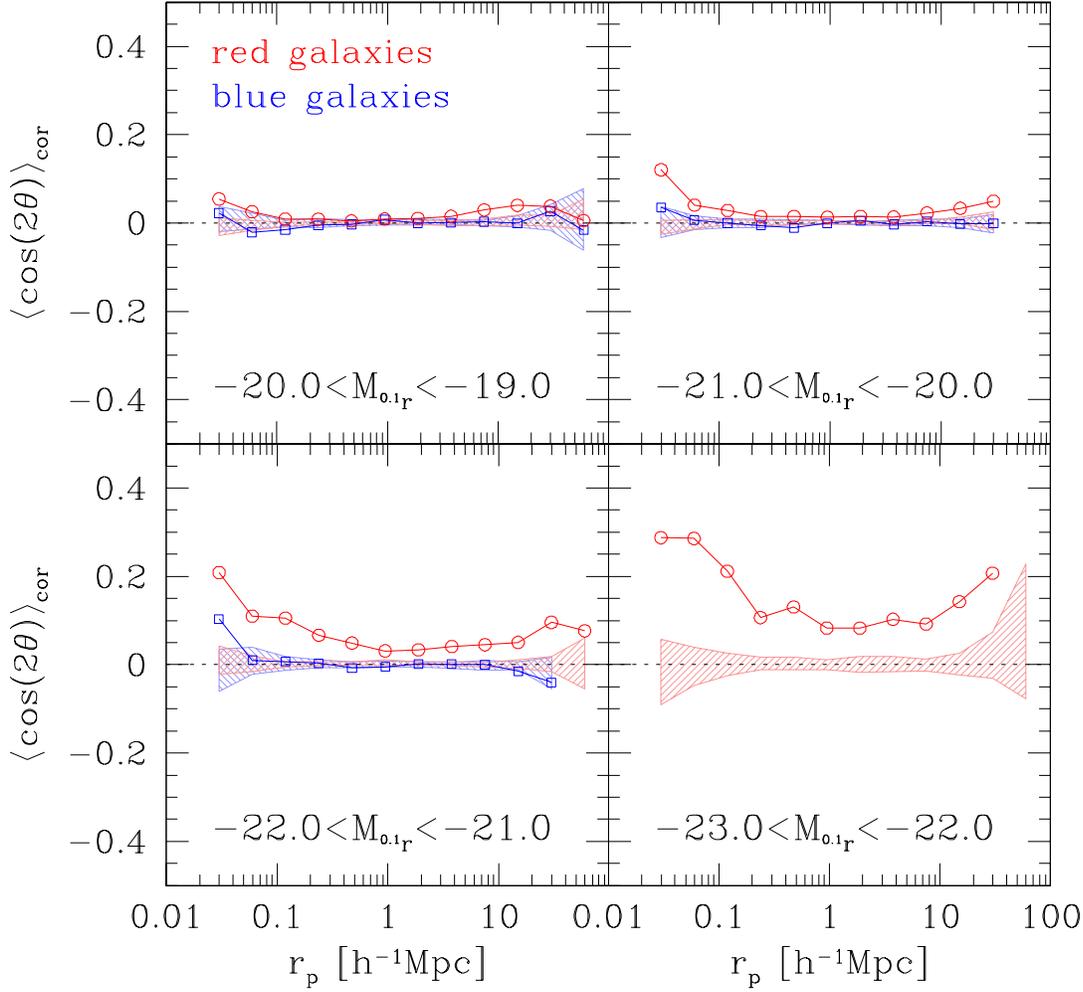}
  \caption{\label{fig:cos2theta} $\cos(2\theta_p)$  statistic for SDSS
    data.  The results for red and blue main galaxies are displayed by
    red  and  blue  lines.   The  galaxies are  subdivided  into  four
    $r$-band absolute magnitude bins as indicated.  The shaded regions
    in  red and  blue show  the $1\sigma$  variance between  10 random
    samples in which the position  angles are shuffled at random among
    the main galaxies.  The reference sample is the same for all these
    measurements  and comprises  all galaxies  with  $r$-band absolute
    magnitudes in $-23<  M_{^{0.1}r} < -17$. No results  are shown for
    blue  galaxies in  the  highest luminosity  since, by  definition,
    they are red galaxies.}
\end{figure}
Fig.~\ref{fig:cos2theta}  displays the  $\cos(2\theta_p)$-statistic as
discussed  in \S~\ref{sec:cos2theta}.   The results  for red  and blue
galaxies are displayed by red and blue lines.  The shaded regions show
the $1\sigma$ variance between the  10 random samples.  The signal for
the red  galaxies shows a  strong dependence on luminosity.   There is
basically no signal for red galaxies in the lowest luminosity bin. For
$r$-band  absolute  magnitudes,   ranging  from  $M_{^{0.1}r}=-20$  to
$M_{^{0.1}r}=-23$  the  signal systematically  lies  above the  shaded
region,  indicating that  it can  not  be explained  by the  intrinsic
scatter inherent to the data.  The signal is significant on all scales
probed  ($\leq60\hMpc$) and  becomes more  pronounced  with increasing
luminosity.  Blue  galaxies do not  show any indication  for alignment
with  the   reference  galaxy  distribution.  Fig.~\ref{fig:cos2theta}
suggests that red  galaxies with $L\gtrsim L_\ast$ tend  to be aligned
with large-scale structure out to  at least $60\hMpc$.  This result is
very similar to that from the alignment correlation function discussed
in  \S~\ref{sec:sdss.ali}.  It  supports the  picture of  a physically
distinct origin for the orientations of red and blue galaxies.
\subsubsection{Alignment of group central galaxies}
\label{sec:sdss.cos.cen}
\begin{figure}
  \includegraphics[width=0.95\textwidth]{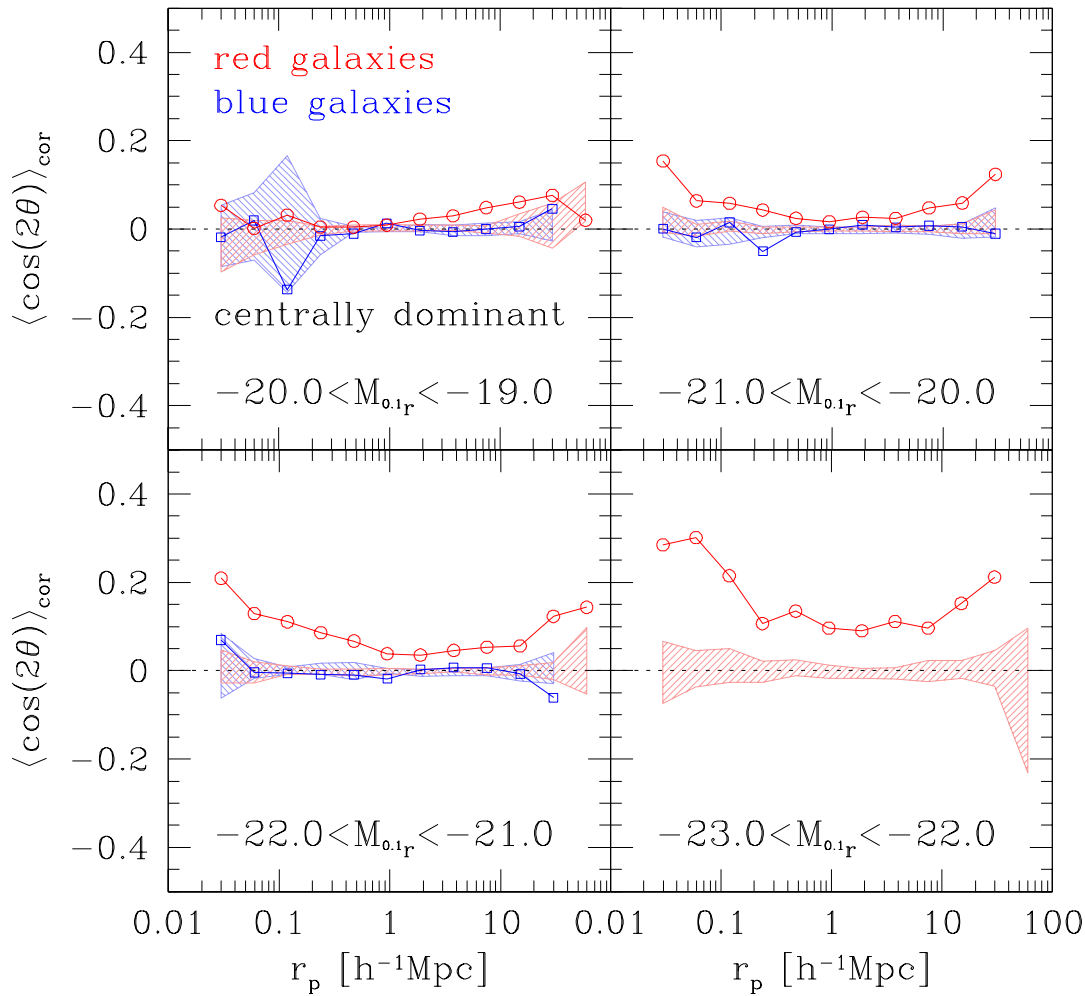}
  \caption{\label{fig:central}  $\cos(2\theta_p)$  statistic for  SDSS
    central   galaxies.   See   the   text  for   our  definition   of
    `central'. Results for red and  blue galaxies are displayed by red
    and  blue   lines.   All   symbols  are  the   same  as   used  in
    Fig.~\ref{fig:cos2theta}.}
\end{figure}
Alignment between clusters of galaxies  has been detected out to large
scales    $\sim100\hMpc$)   \citep[e.g.,][]{Binggeli-82,   Plionis-94,
  Hashimoto-Henry-Bohringer-07}.   In  addition  there  exists  strong
evidence that  the orientations of  central galaxies are  aligned with
the orientations of their parent groups or clusters.  Now the question
arises whether  an exclusion of  non-central galaxies may  enhance the
alignment  signal measured  before.  To  this  end we  redo the  above
analysis  excluding {\it  non-central} galaxies.   This  approach also
facilitates a  subsequent comparison with N-body  results, since there
only  the   orientations  of  central  galaxies   are  available  (see
\S~\ref{sec:mille.cos}).

Central galaxies are found in the following way. For each main galaxy,
we  use its  $r$-band absolute  magnitude  to calculate  a halo  mass,
according to  the relation between central galaxy  luminosity and halo
mass as given  by \cite{Yang-05b}. We then estimate  a `virial' radius
using the  model of \cite{Eke-Navarro-Steinmetz-01}.  We  find all the
companions around  the galaxy within this  radius and a  line of sight
velocity separation of $\pm3000\kms$ and compare the luminosity of the
galaxy to that of the companions.   If the galaxy is brighter than any
of its companions and it is not inside the virial radius of a farther,
brighter galaxy it is classified as a `central' galaxy.  Otherwise, it
is a  `non-central' galaxy.  This  method will inevitably  mis-judge a
certain fraction of  galaxies to be non-central due  to the appearance
of  more luminous  interlopers  \citep[cf.][]{Yang-05a}.  However,  it
should  be sufficient  to estimate  the signal  for  central galaxies.
This can  then be compared  to the signal  for the complete  sample as
well as to simulations.

Fig.~\ref{fig:central} displays the alignment signals based on central
galaxies only.  As before we do not find any alignment signal for blue
galaxies. The upper two luminosity bins show only marginal differences
when compared  to the  complete sample. This  can be explained  by the
fact  that  most bright  galaxies  are  central.   For the  lower  two
luminosity bins the alignment signal increases by a few per cent which
indicates a certain reduction of the alignment signal for the complete
sample which includes non-central galaxies (Fig.~\ref{fig:cos2theta}).
We come  to the conclusion that  the alignment signal  is not strongly
enhanced by the exclusion of non-central galaxies which is due in part
to the fact that luminous galaxies usually are central.
\section{Comparison to the alignment of MS galaxies}
\label{sec:mille}
In  the second  part of  this study  we apply  our new  analysis tools
(alignment  correlation  function  and  $\cos(2\theta$)-statistic)  to
semi-analytic galaxies in the MS.  Orientations have to be assigned to
the model galaxies and  we investigate different approaches, comparing
the resulting alignment signals with  the observed ones.  Since we did
not  detect any  alignment signal  for blue  galaxies we  restrict our
orientation assignment  to red  model galaxies.  This  restriction may
also be justified by a  slightly different kind of reasoning.  Namely,
red  luminous galaxies are  predominantly elliptical  galaxies.  Those
are thought to form via merging of proto-galaxies which is basically a
collisionless process.  Thus N-body simulations may be able to recover
some  characteristic features  of  red galaxies,  in particular  their
orientation.   Blue galaxies,  on the  other side,  are  mainly spiral
galaxies  with their  properties heavily  dependent on  baryonic, i.e.
collisional,  physics.   Thus  their  properties may  be  only  poorly
recovered  by N-body  simulations.  This  picture may  change  at high
redshifts. At  that time, also  the orientation of many  blue galaxies
could be a product of gas-rich mergers.

In the  following two paragraphs  we list some  details of the  MS and
describe    how    we   estimate    orientations    for   the    model
galaxies.  Subsequently we redo  the analysis  already carried  out on
SDSS galaxy samples and we compare the results.
\subsection{The simulation and the galaxy sample}
\label{sec:mille.details}
The  Millennium  Simulation  \citep{Springel-05a} adopted  concordance
values  for  the parameters  of  a  flat  $\Lambda$ cold  dark  matter
($\Lambda$CDM)  cosmological  model,   $\Omega_{\rm  dm}=  0.205$  and
$\Omega_{\rm b}= 0.045$ for the  current densities in CDM and baryons,
$h= 0.73$ for the present  dimensionless value of the Hubble constant,
$\sigma_8= 0.9$  for the  rms linear mass  fluctuation in a  sphere of
radius $8 \hMpc$ Mpc extrapolated to  $z= 0$, and $n= 1$ for the slope
of  the  primordial  fluctuation  spectrum.  The  simulation  followed
$2160^3$ dark matter particles from z= 127 to the present day within a
cubic  region $500\hMpc$ on  a side  resulting in  individual particle
masses  of  $8.6\times10^8\hMsol$.   The  gravitational  force  had  a
Plummer-equivalent  comoving  softening   of  $5\hkpc$.   The  Tree-PM
$N$-body code  GADGET2 \citep{Springel-05b} was used to  carry out the
simulation and the full data were stored 64 times spaced approximately
equally in  the logarithm of  the expansion factor.   This information
makes  it possible  to construct  trees that  store  detailed assembly
histories for each dark matter halo present at z= 0.

The halos  are found by a  two-step procedure.  In the  first step all
collapsed  halos with  at least  20 particles  are identified  using a
friends-of-friends (FoF) group-finder with  linking parameter b = 0.2.
These  objects   will  be  referred  to  as   {\it  FoF-halos}.   Then
post-processing    with    the    substructure    algorithm    SUBFIND
\citep{Springel-01} subdivides each FoF-halo  into a set of self-bound
      {\it sub-halos}.  Here we  only consider the {\it main sub-halo}
      which is the sub-halo with the most massive progenitor among all
      sub-halos  belonging to  the same  FoF-halo.  We  refer  to this
      sub-halo  as the  parent  sub-halo associated  with the  central
      galaxy.

Based on  the assembly histories  individual halos are  populated with
semi-analytic  galaxies  for which  many  `observable' quantities  are
generated.   For a  detailed description  of the  construction  of the
semi-analytic galaxy  catalog we refer the  reader to \cite{Croton-06}
and \cite{DeLucia-Blaizot-07}. Here we use the DeLucia2006a\_SDSS2MASS
catalog  ({\tt   http://www.g-vo.org/MyMillennium2/})  which  provides
synthetic magnitudes  through SDSS filters,  alternatively to classify
the model galaxies as well as  to tag their parent sub-halos.  The use
of semi-analytic  galaxies facilitates comparison  to the observations
and is also an elegant way to characterize the assembly history of the
host halo by a few numbers.

\begin{table}
  \begin{center}
    \begin{tabular}{c||c|c|c|c}
      $M_r$&$[-20,-19]$&$[-21,-20]$&$[-22,-21]$&$[-23,-22]$\\\hline\hline
      $N_{\rm halo}$           &16027  &98618  &167513 &53065\\\hline
      $\lra{N_{\rm sub}}$      &489    &1672   &4707   &18662\\\hline
      $\lra{N_{\rm cen}}$      &31     &83     &164    &278
    \end{tabular}
  \end{center}
  \caption{\label{tab:misali}  Particle  number  statistics  for  host
    halos }
\end{table}
Our  standard  reference sample  comprises  a  random subset  (1103640
galaxies)  of  all   semi-analytic  galaxies  with  $r$-band  absolute
magnitudes of $-23  \leq M_r\leq -17$ (in total  11027979 galaxies) at
$z=0$. This standard  reference sample is used for  all the subsequent
analysis based on MS galaxies.  The parent galaxy sample comprises all
red, central galaxies within the same magnitude range as the reference
galaxies ( $-23 \leq M_r\leq  -17$).  A galaxy is red if $g-r\geq0.7$,
where $g$ and  $r$ are synthetic magnitudes in  the corresponding SDSS
bands.  A  galaxy is central if it  is the dominant galaxy  in a given
FoF-halo.  Expressed  in the  MS terminology it  is the (only)  type 0
galaxy  within this  FoF-halo.   This definition  comes  close to  the
selection  we  have  employed  to  identify central  galaxies  in  the
observational analysis  above.  In analogy  to the SDSS sample  the MS
galaxies  are  split into  four  $r$-band  magnitude  bins with  equal
intervals of  one magnitude.  Each  model galaxy is associated  with a
dark matter  halo which will be  used to determine  the orientation of
the galaxy.  For the four magnitude bins Table~\ref{tab:misali} lists:
the number of halos ($N_{\rm  halo}$); the average number of particles
belonging to these halos,  $\lra{N_{\rm sub}}$; and the average number
of particles belonging to the  central part of the halos, $\lra{N_{\rm
    cen}}$. The  determination of the  number of central  particles is
related to  the computation  of the central  orientations and  will be
discussed  below.   We conclude  this  paragraph  by emphasizing  that
orientations are  only determined  for red, central  galaxies, because
only for those  galaxies do we expect a  tight correlation between the
orientations of galaxy and halo.
\subsection{The orientations of semi-analytic galaxies}
\label{sec:mille.ori}
As a proxy for the major axis of the central galaxy we adopt the major
axis of the projected moment  of inertia tensor of the parent sub-halo
\citep[cf.][]{Agustsson-Brainerd-06b,Kang-07,Faltenbacher-08,Knebe-08a,
  Okumura-Jing-Li-08}.  Since  real galaxies are  seen in projection,
we  project  the dark  matter  distribution  before diagonalizing  the
moment  of  inertia tensor.  We  determine  the  orientations of  each
central  galaxy in  two  alternative  ways.  We  use  the dark  matter
distribution of  the parent sub-halo,  (i.e. the main sub-halo  of the
FoF-halo)  or  we  only  use  its  central  part  to  approximate  the
orientation of the  central galaxy. The former is  done to compare the
results  to earlier  work  while  the latter  better  mimics the  true
physical  circumstance \citep[cf.][]{Faltenbacher-08,  Knebe-08a}.  We
will refer to  the former as {\it halo orientation}  and the latter as
{\it central} or {\it galaxy orientation}.  The central orientation is
determined in  an iterative  way.  We begin  by computing a  number of
central particles $N_{cen}$ using the following formula
\begin{equation}
\label{eqn:cooray}
  N_{cen} = 
  \begin{cases}
    37\ &\times\ N_{sub}^{0.2}\ \text{ for } N_{sub}\geq 4900 \\
    0.225 &\times\ N_{sub}^{0.8}\ \text{ for } N_{sub}< 4900 
  \end{cases}
\end{equation}
where $N_{sub}$ is the number of particles within the parent sub-halo.
This formula is a two-power-law approximation to the observed relation
between the  luminosity of a central  galaxy and its  parent halo mass
\citep[e.g.][]{Cooray-Milosavljevic-05}.   It reflects  the assumption
that  the  distribution  of   a  galaxy's  stellar  component  can  be
represented by  the central matter distribution of  its simulated dark
matter halo.  The mass representing  the galaxy then has to follow the
observed  scaling  laws.  Now,  let  $r_{init}$  be  the radius  which
encloses  the $N_{cen}$.   In a  first  iteration step  the moment  of
inertia is computed within  the $r_{init}$.  The resulting axis ratios
are used  to cut  out a  central ellipsoidal where  the length  of the
intermediate axis is fixed to  $r_{init}$. Based on this new subset of
particles the  moment of inertia  is computed anew.  The  iteration is
continued  until  the   central  orientation  converges.   Fixing  the
intermediate axis results in approximately unchanging particle numbers
within consecutive ellipsoids.  Up to this point the computations were
done in  three dimensions.   In a last  step all particles  within the
final ellipsoid are projected onto  the plane of the sky.  The inertia
tensor of this 2-D distribution determines the central orientation.

\begin{figure}
  \includegraphics[width=0.95\textwidth]{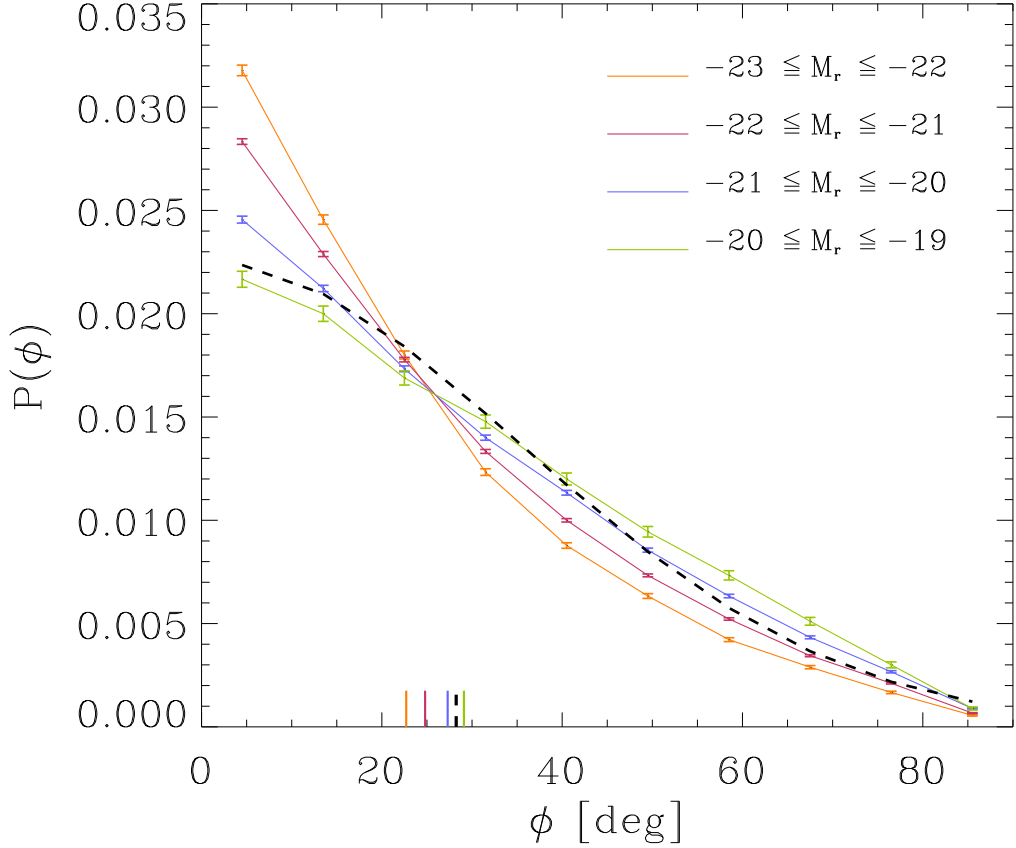}
  \caption{\label{fig:misali}   Probability    distribution   of   the
    misalignment angle between  halo and central (galaxy) orientations
    for  a series of  magnitude bins  as indicated.   Errors represent
    Poisson uncertainties. The small vertical lines on the bottom axis
    denote  the mean  values of  the distributions.   The  dashed line
    display the  result of  \cite{Okumura-Jing-Li-08} and its  mean we
    show as black dashed mark on the bottom axis.}
\end{figure}
In  Fig.~\ref{fig:misali}  the  difference  between halo  and  central
(galaxy) orientations are  illustrated by the probability distribution
function  (PDF) of  the  misalignment angle  between  them.  The  mean
values of the misalignment for each magnitude bin are indicated by the
little  vertical   marks  on  the   bottom  axis.   As   indicated  in
Table~\ref{tab:misali}, on average there are only 31 particles used to
determine the  central orientations within the  lowest luminosity bin.
Thus, Poisson  noise adds  an additional misalignment  component which
contributes to  the flattening  of the PDF  at small angles.   In fact
calculating the  misalignment using central volumes twice  as large as
given by Eq.~\ref{eqn:cooray}  substantially reduces the flattening in
the lowest luminosity  bin. However, since we aim  to probe the volume
actually occupied by  the stellar component of the  galaxies we adhere
to Eq.~\ref{eqn:cooray} keeping in  mind that the central orientations
for the low luminosity galaxies are poorly resolved.  Such effects are
less  important  for the  higher  luminosity  bins.   The two  highest
luminosity  bins use  160 and  270 particles  on average  which should
result    in   a    robust   determination    of    the   orientations
\citep[cf.][]{Jing-02}.  For  comparison we also  display the Gaussian
PDF   suggested  by  \cite{Okumura-Jing-Li-08},   who  found   a  mean
misalignment angle  of $\sim25^\circ$ between the halo  and the galaxy
orientation. Comparing the orientations  of central galaxies and their
host   halos    a   similar    misalignment   has   been    found   by
\cite{Agustsson-Brainerd-06b}  and   \cite{Kang-07}.   Somewhat  lower
values are quoted in \cite{Wang-08}. The misalignment weakens slightly
with increasing luminosity  of the central galaxy, though  this may be
due to the resolution effects discussed above.
\subsection{Alignment correlation function}
\label{sec:mille.ali}
With  these  orientations  and  the model  magnitudes  the  alignment,
correlation function can  be computed for the galaxies  in the MS.  We
will adopt the same luminosity bins  used for our analysis of the SDSS
data, simplifying the comparison between observations and simulations.
For  the computation  of  the  alignment angle  we  apply the  distant
observer approximation.  To that end the z-direction in the simulation
is  chosen  to  be parallel  to  the  line  of  sight.  To  mimic  the
observational approach the maximum projected separation of pairs along
the line of sight is $\pi_{\rm max}=40\hMpc$.

\begin{figure}
  \includegraphics[width=0.95\textwidth]{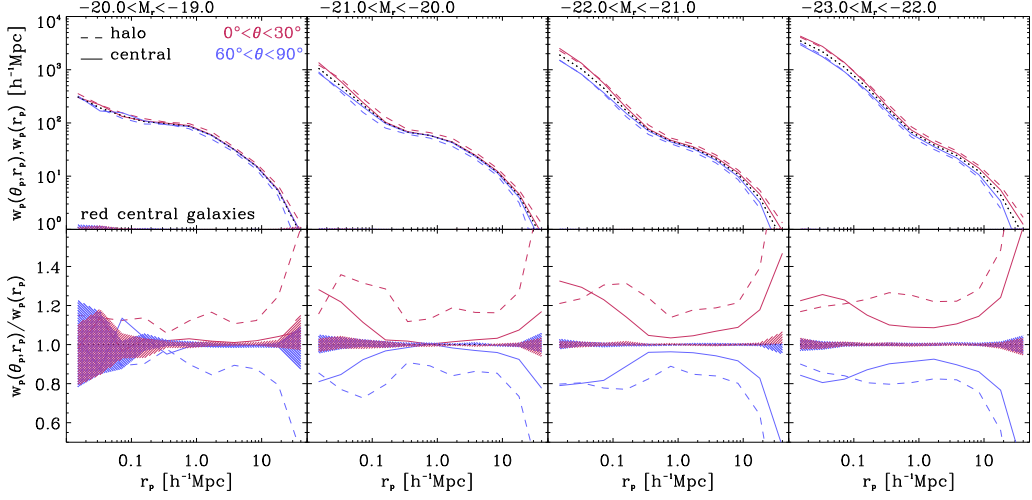}
  \caption{\label{fig:C33} {\it Upper  panels:} Projected MS alignment
    correlation   function,  $w_p(\theta_p,r_p)$,   for   red  central
    galaxies  in different  intervals of  $r$-band  absolute magnitude
    using the standard reference  sample. The colors correspond to the
    sectors as  indicated.  The  solid and dashed  lines are  based on
    central  and halo  orientations.  The  dotted line  represents the
    traditional   projected   two-point  cross-correlation   function,
    $w_p(r_p)$.  {\it  Lower panels:} Ratio between  the alignment and
    traditional correlation functions.  The  color code is the same as
    used above.   Red and blue shaded regions  indicate the $1-\sigma$
    variance based on 10 samples in which the central orientations are
    shuffled at random among the galaxies.}
\end{figure}
Fig.~\ref{fig:C33} shows  the alignment correlation  function for four
luminosity  bins  which  can  be   compared  to  the  upper  panel  in
Fig.~\ref{fig:wrp}. Also the luminosity  range of the reference sample
is chosen  to be the same  as in our observational  analysis.  At this
point we would like to emphasize that the MS main sample only includes
central galaxies.   This causes the  pronounced bend at a  few hundred
$\hkpc$ corresponding roughly to virial  radii of the host halos.  The
general  behavior,   however,  is  very   similar  to  that   seen  in
Fig.~\ref{fig:wrp}.   The correlation function  is enhanced  along the
major  axes  of  the  red galaxies  ($0^\circ<\theta_p<30^\circ$)  and
reduced   perpendicular   to  them.    For   galaxies  brighter   than
$M_r\leq-20$ a systematic alignment/anti-alignment signal is seen over
the  entire  separation range  probed  (out  to  $60\hMpc$).  And  the
strength of the alignment effect  increases with the luminosity of the
galaxies.

In addition  we find that,  except for small (intra-halo)  scales, the
alignment  signal based on  the halo  orientations is  more pronounced
than  that based  on the  central (galaxy)  orientations. This  can be
explained by  the misalignment  between halo and  central orientations
discussed  in  \S~\ref{sec:mille.ori}.  A  comparison  with the  upper
panel of  Fig.~\ref{fig:wrp} reveals that the alignment  based on halo
orientations overestimates the observational signal whereas that based
on  the central  orientations reproduces  the amplitudes  obtained for
SDSS galaxies quite well.

\begin{figure}
  \centering
  \includegraphics[width=0.38\textwidth,clip=true]{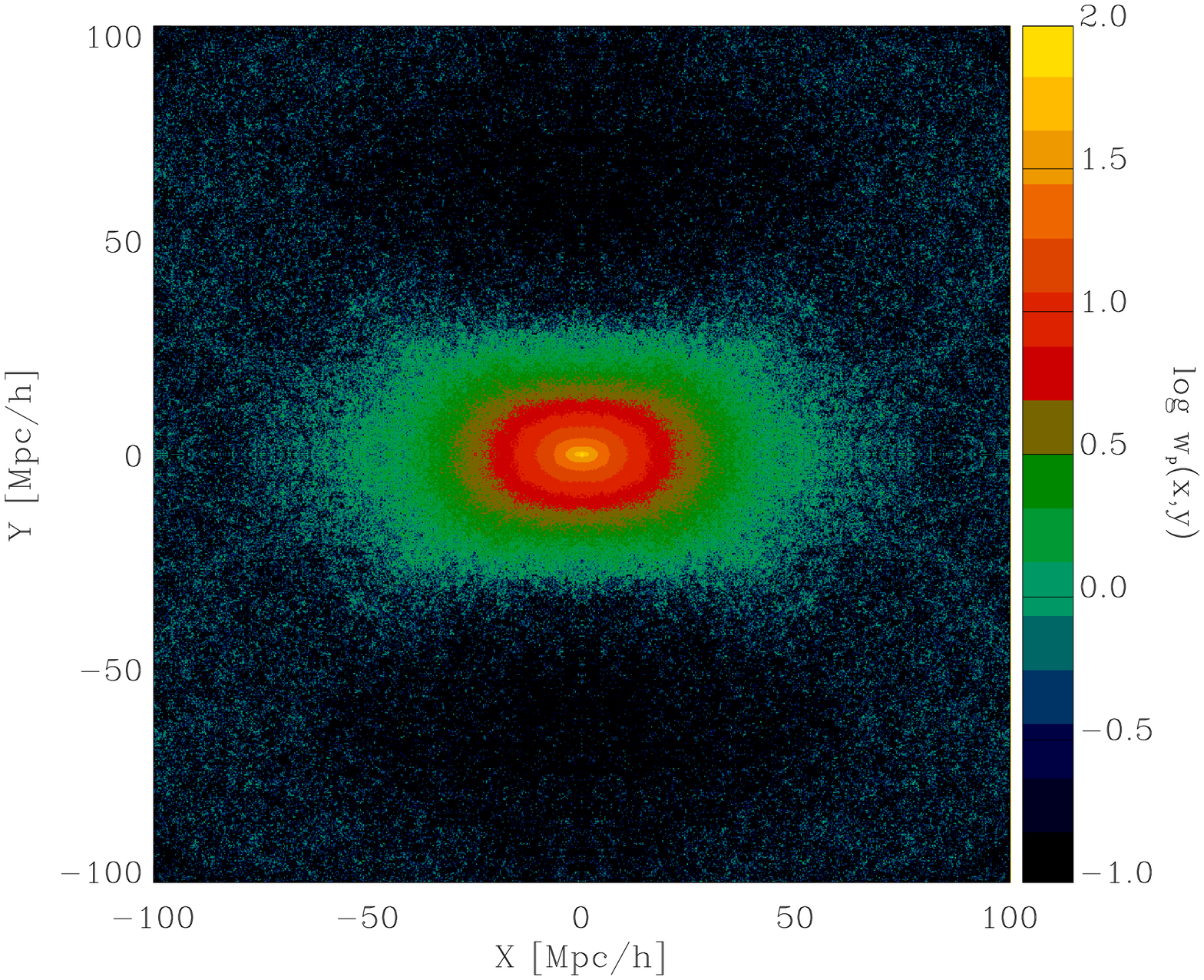}
  \hspace{0.08\textwidth}
  \includegraphics[width=0.38\textwidth,clip=true]{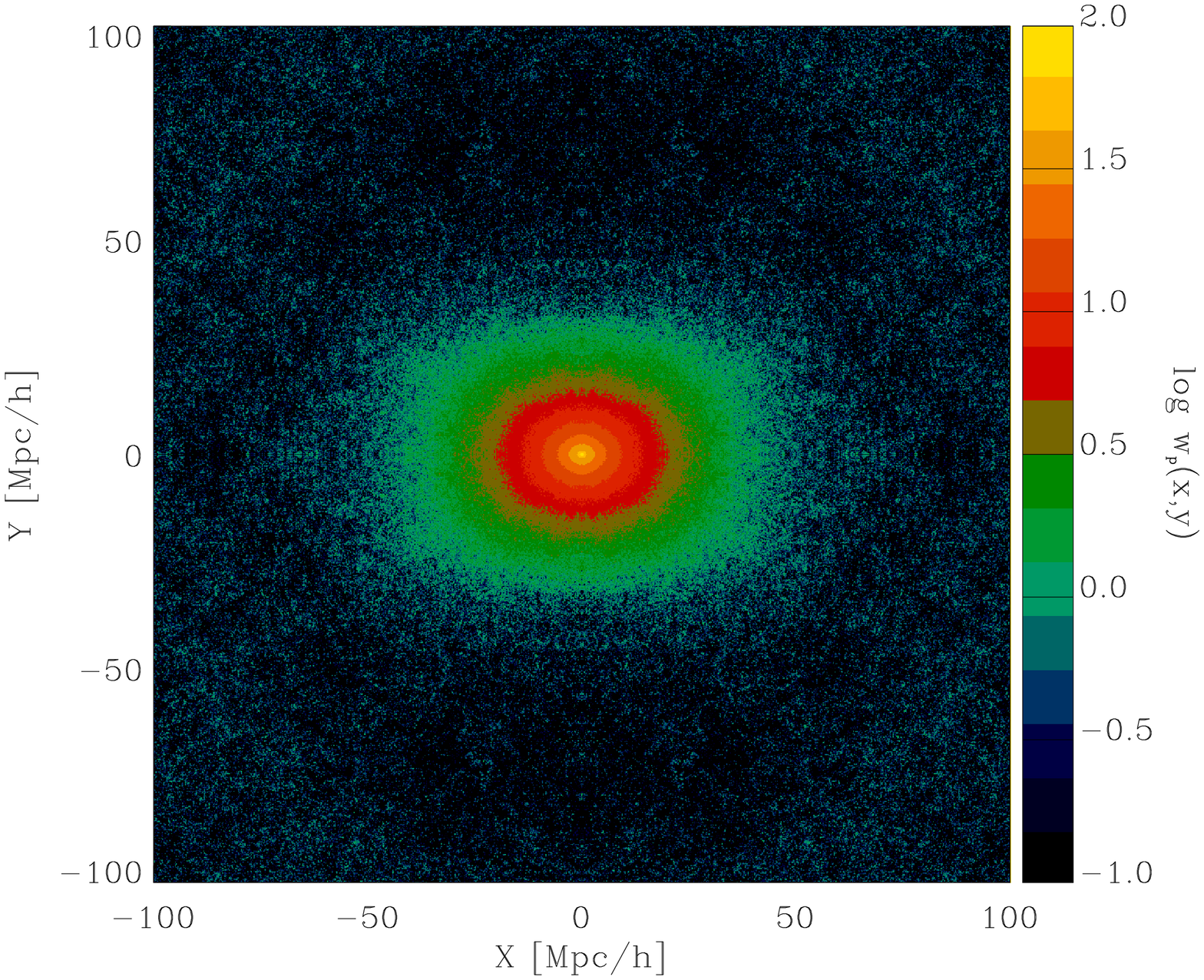}\\
  \includegraphics[width=0.38\textwidth,clip=true]{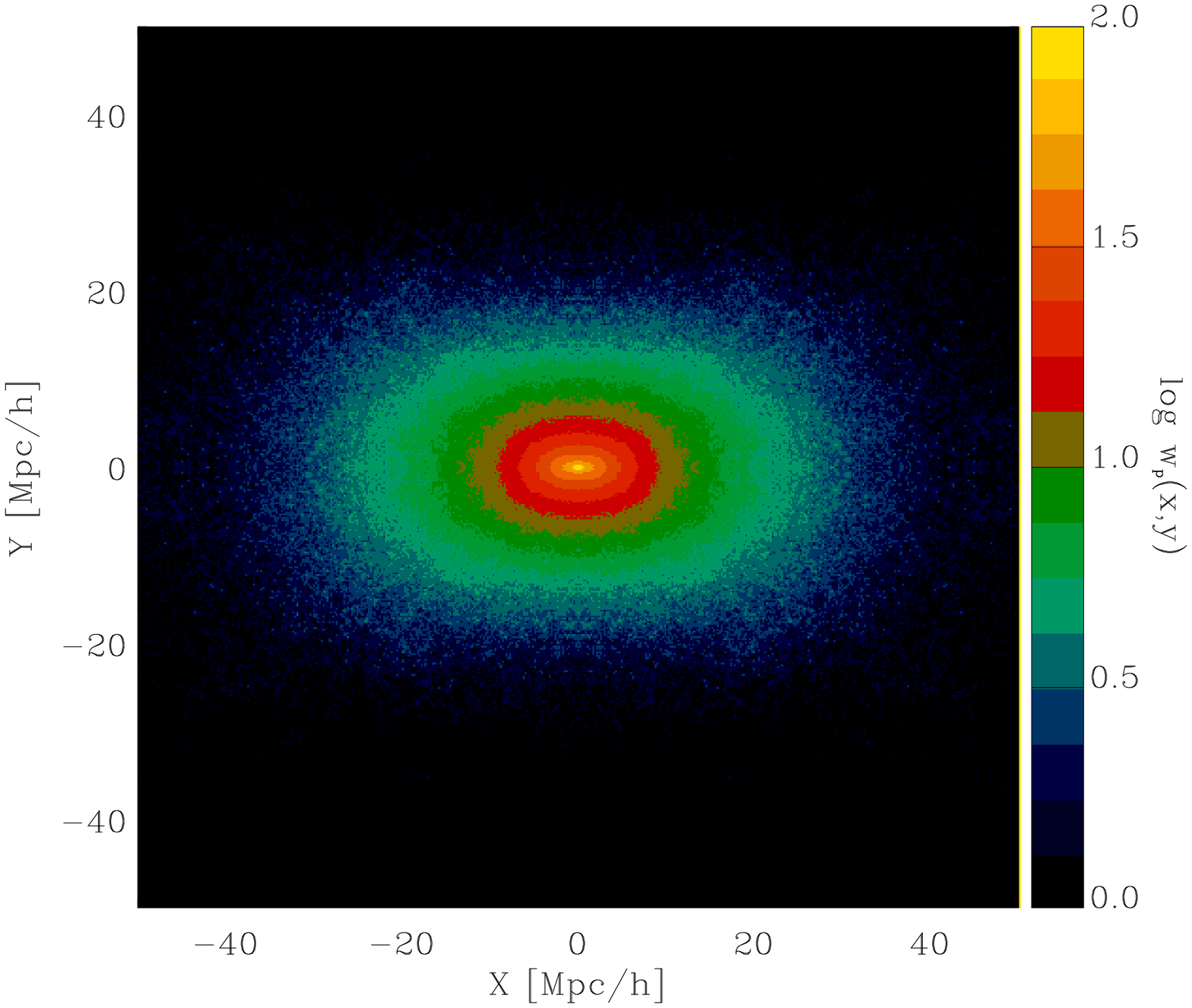}
  \hspace{0.08\textwidth}
  \includegraphics[width=0.38\textwidth,clip=true]{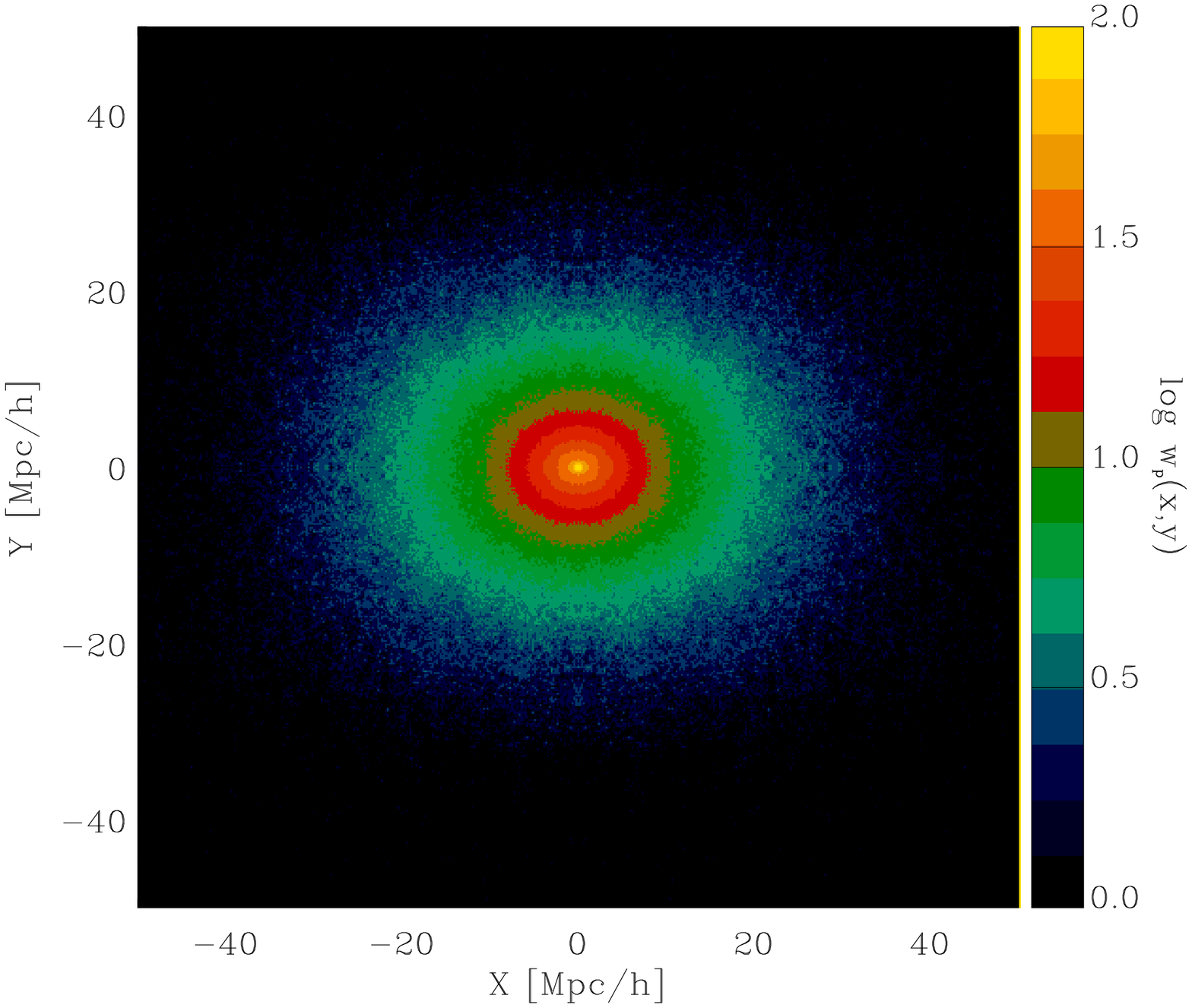}\\
  \includegraphics[width=0.38\textwidth,clip=true]{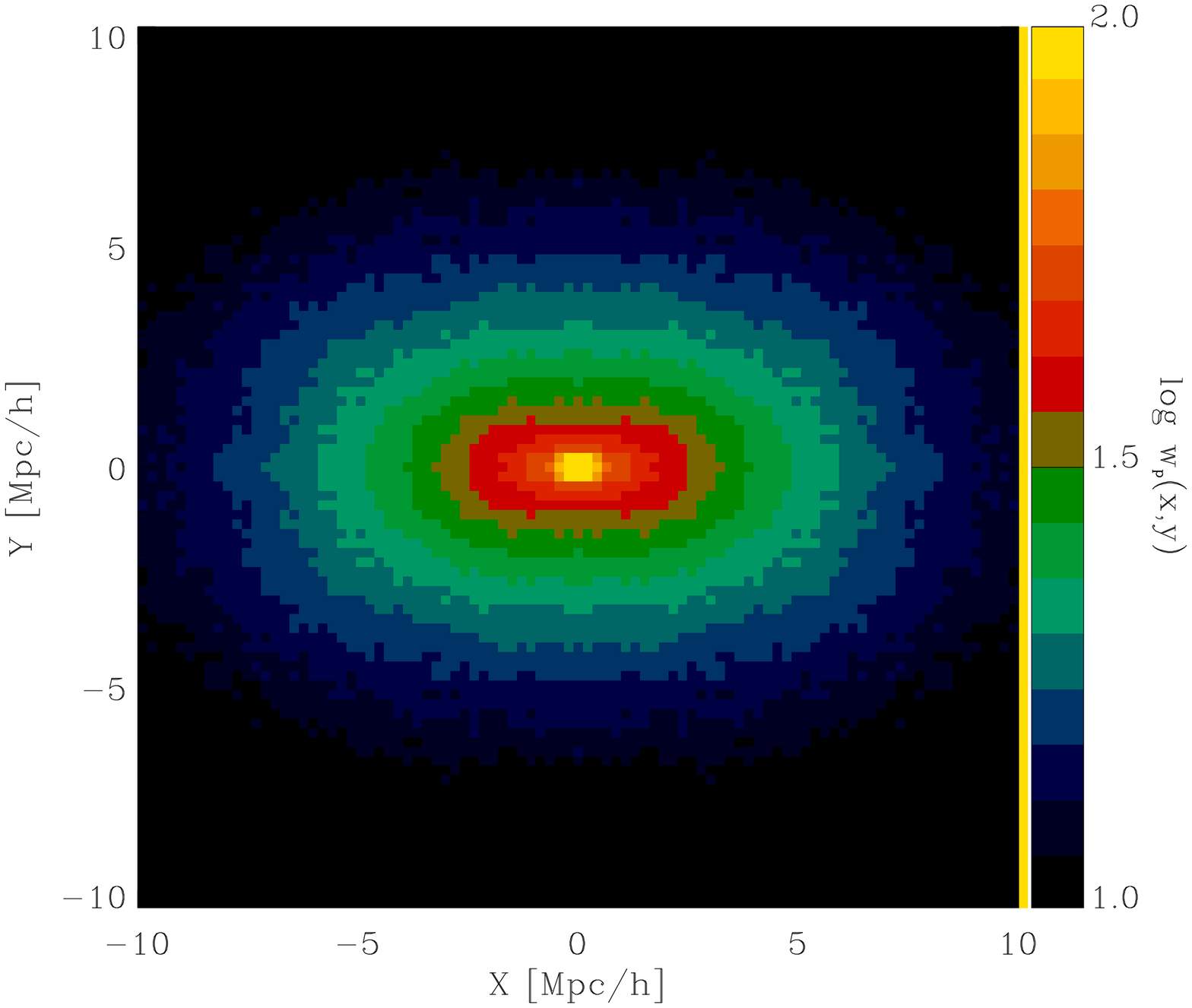}
  \hspace{0.08\textwidth}
  \includegraphics[width=0.38\textwidth,clip=true]{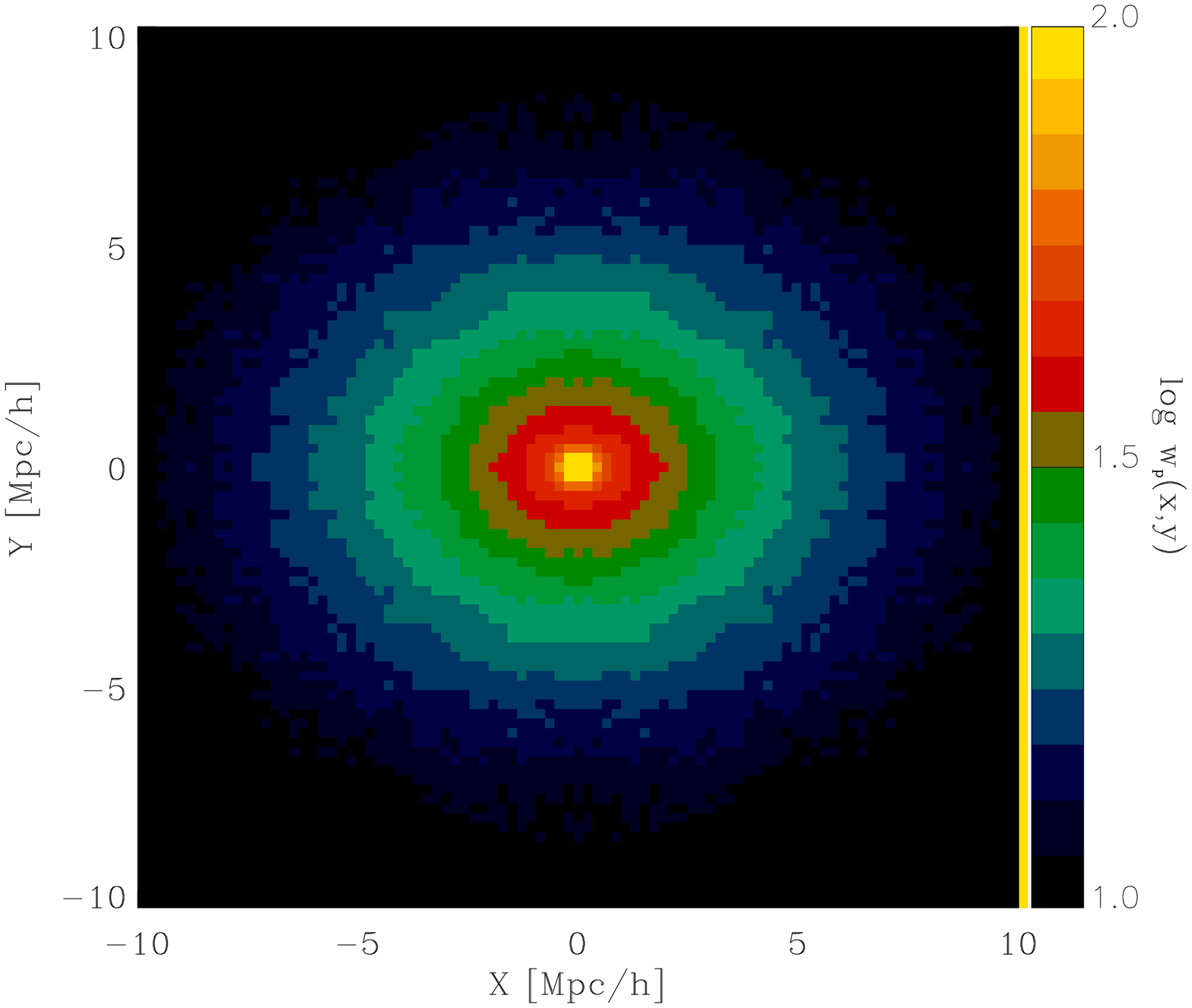}
  \caption{\label{fig:wxy} 2D alignment  correlation function based on
    the DeLucia2006a\_SDSS2MASS  semi-analytic galaxy catalog  for the
    MS.   The  main sample  consists  of  red,  central galaxies  with
    absolute  $r$-band  magnitudes   between  $-21$  and  $-22$.   The
    orientations of  the galaxies are  aligned with the  x-axis.  Left
    and  right  panels  display  results  based on  halo  and  central
    orientations.    Panels  from   top  to   bottom  show   the  same
    distribution for  different scales.  The color  code indicates the
    correlation  amplitude,  its range  changes  from  top to  bottom.
    Pixels  with  values  below  and  above the  indicated  range  are
    displayed black and yellow,  respectively.  The ratios of minor to
    major axes  at $\log w_p(x,y)=1.5$,  $1$ and $0.5$ are  0.57, 0.61
    and  0.42  for the  halo  and 0.75,  0.80,  0.75  for the  central
    orientations.}
\end{figure}

So far we have restricted the computation of the alignment correlation
to 3 angular bins, a relatively coarse segmentation.  This facilitates
the direct  comparison with the  SDSS results, derived above,  where a
less coarse binning would  result in excessive Poisson noise.  However
the  MS   provides  much   better  statistics,  making   possible  the
computation  of  a  2D  alignment  correlation  map  as  displayed  in
Fig.~\ref{fig:wxy}.    This  shows  the   alignment  cross-correlation
between a sample consisting of  red ($g-r \geq 0.7$), central galaxies
with  $r$-band  luminosities between  -21  and  -22  and our  standard
reference sample.   The left  and right panels  are based on  halo and
central orientations,  respectively.  The  panels in a  column display
consecutive blowups of the same map.  The correlation amplitude can be
read off using the adjacent  color bars.  Note, the range changes from
top  to  bottom.   The  ratios  of  minor  to  major  axes  for  $\log
w_p(x,y)=1.5$,  $1$ and $0.5$  are 0.57,  0.61 and  0.42 for  the halo
orientations and  0.75, 0.80, 0.75  for the central  orientations.  In
agreement with  the results above, the alignment  correlation based on
halo  orientations  is more  anisotropic.   It  is  striking that  the
flattening of the  counts is almost constant over  the range of scales
we can measure.

The  main   conclusion  to  be  drawn   from  Figs.~\ref{fig:C33}  and
\ref{fig:wxy} is that  the halos of red luminous  galaxies are aligned
with the  surrounding galaxy distribution  out to at  least $60\hMpc$.
If  the   orientations  of  the  galaxies  are   approximated  by  the
orientation of the  halos as a whole, the  alignment signal is strong.
The misalignment between the halo  and central region, as discussed in
\S~\ref{sec:mille.ori},  causes a weakening  of the  correlation.  The
detection of  the alignment between  red galaxies and  the large-scale
structure provides  a tool  to test predictions  for the  alignment of
galaxies with their halos.
\subsection{$\cos(2\theta)$-statistic}
\label{sec:mille.cos}
\begin{figure}[t]
  \centering
  \includegraphics[width=0.95\textwidth]{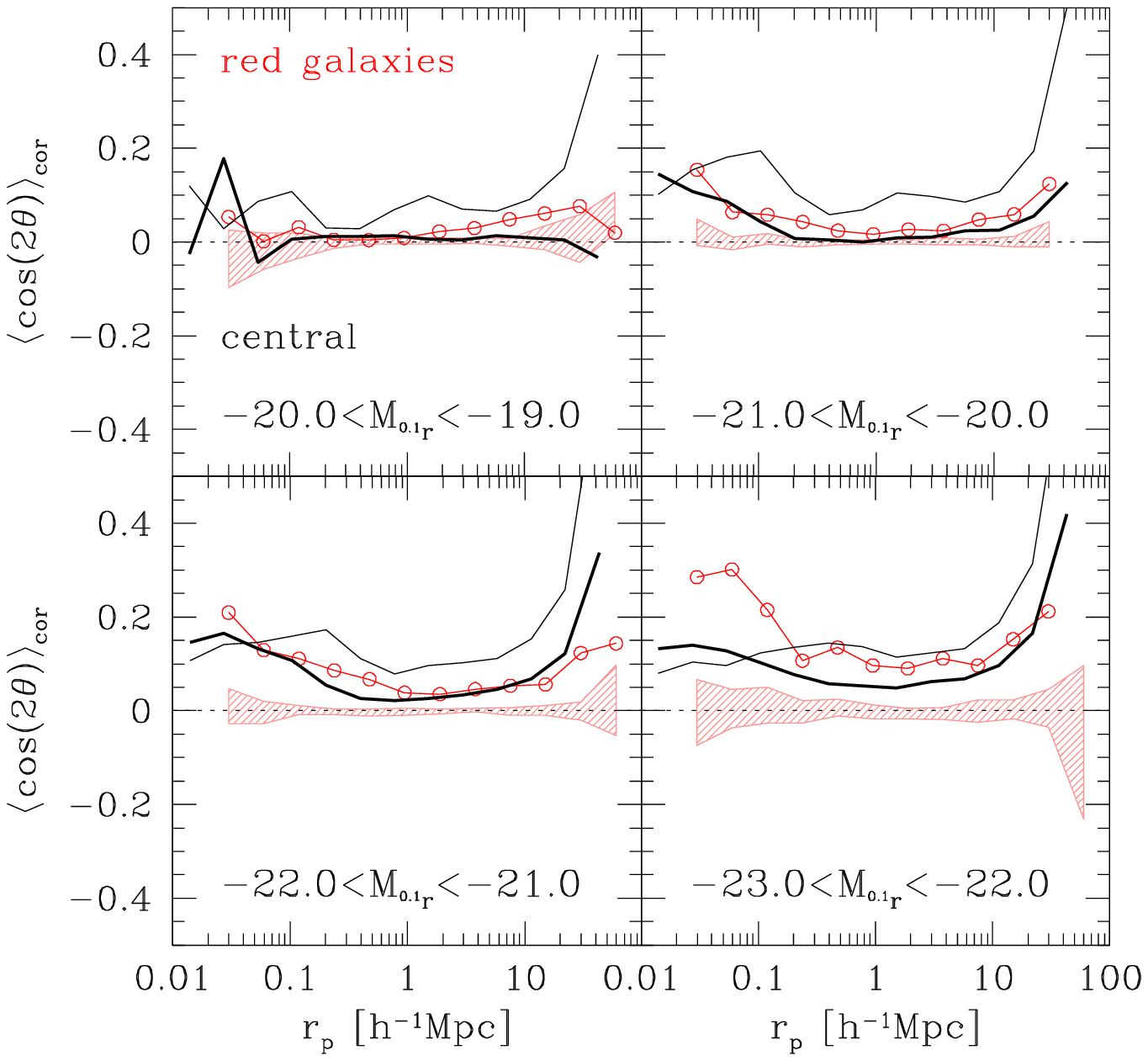}
  \caption{\label{fig:comparison}  Comparison  of  the  $\cos(2\theta)$
    signal for  red SDSS  galaxies and semi-analytic  galaxies.  The
    red lines and symbols  are copied from Fig.~\ref{fig:central}. The
    black  lines  display   the  corresponding  results  derived  from
    semi-analytic  galaxy catalogs.   Thin  lines are  based on  the
    halo-orientations and thick lines represent results
    based model central-orientations.  }
\end{figure}
To    facilitate   the    comparison   of    observed    and   modeled
$\cos(2\theta)$-statistics   we  follow  the   observational  approach
discussed  in   \S~\ref{sec:sdss.ali},  i.e  we   use  the  equivalent
luminosity  and color  cuts.  Again  the  analysis is  carried out  in
projected  space where the  z-axis is  chosen to  lie parallel  to the
line-of-sight and the distant  observer approximation is applied.  The
maximum   line-of-sight  separation   between   correlated  pairs   is
restricted to $40\hMpc$. Only  the signal based on red ($g-r\geq0.7$),
central galaxies  is compared to the corresponding  SDSS results.  The
standard MS reference sample is used.

In  Fig.~\ref{fig:comparison}  the   results  for  central,  red  SDSS
galaxies are copied  from Fig.~\ref{fig:central}, including the shaded
region  reflecting the  intrinsic scatter  in the  observational data.
The black lines display the corresponding signals derived from the MS.
Thick  lines  represent the  results  based  on central  orientations,
whereas  thin lines  are based  on halo  orientations.  In  the lowest
luminosity bin the halo orientations result in a substantial alignment
signal whereas  the signal based  on the central orientations  and the
SDSS data are in agreement with no detected alignment signal.  For the
next  two luminosity  bins the  signal based  on  central orientations
agrees  well  with  the  SDSS  data.  The  mismatch  for  the  largest
separation bin may be associated with zero-crossing of the correlation
function,  because the  response of  the  $\cos(2\theta)$-statistic is
very sensitive near  this point.  In the highest  luminosity bin there
appears  a  disagreement  between  the  MS  signal  based  on  central
orientations for  separations $\leq5\hMpc$ and  the SDSS measurements.
The agreement  is better with  the MS results  based on the halo  as a
whole.  On  larger scales $5\hMpc\lesssim r_p  \lesssim 60\hMpc$ again
there is good agreement with  the signal based on central orientations
whereas that based on overall  halo orientations is an overestimate by
50-100 per cent.

We  are led  to the  following  conclusions: 1)  the alignment  signal
depends on the magnitude of the  central galaxy and/or the mass of the
host halo.   2) The alignment signal  based on the  orientation of the
halo exceeds the  observed signal, in most cases,  whereas the central
orientations generally  result in good  agreement between observations
and  simulations.  3)  The  alignment for  red  central galaxies  with
$M_{^{0.1}r}<-20$    persists   to    the   largest    scales   probed
($\sim60\hMpc$).
\section{Summary}
\label{sec:sum}
In the first part of this study  we use the SDSS DR6 galaxy catalog to
quantify the alignment between galaxies and the large-scale structure.
We develop  two new statistics, namely the  {\it alignment correlation
  function} and the {\it  $\cos(2\theta)$-statistic}.  The former is a
two-dimensional  extension of  the  traditional two-point  correlation
function taking  into account the  orientations of the  main galaxies.
The latter is related to  the ellipticity correlation function used to
analyze  cosmic shear  observations. In  both cases  we  represent the
large-scale  structure  by a  reference  galaxy  sample with  absolute
$r$-band magnitudes between $-17$ and $-23$.

The alignment correlation  function is defined on the  2D plane of the
sky. In our  SDSS analysis we use a polar  coordinate system and probe
the correlation function  along the galaxy major axes  focusing on the
$0^\circ<\theta_p<30^\circ$  sector,  where  $\theta_p$ is  the  angle
between the major axis and  the connecting line to a reference galaxy.
The abundance of reference galaxies perpendicular to the major axes is
represented  by  the  counts within  the  $60^\circ<\theta_p<90^\circ$
sector.  The  main  results  revealed  by  the  alignment  correlation
function are: 1) The major axes of red galaxies with $M_{^{0.1}r}<-20$
are significantly aligned with large-scale structure; 2) The signal is
visible out  to $60\hMpc$ (the  range probed here); 3)  Independent of
luminosity blue galaxies are not aligned with large-scale structure.

The  $\cos(2\theta)$-statistic  also  probes  the  anisotropy  of  the
reference galaxies around  main galaxies and is related  to the galaxy
--- intrinsic   shear    correlation   function   as    discussed   in
\cite{Mandelbaum-06a}  and  \cite{Hirata-07}.    In  contrast  to  the
alignment correlation  function it  does not require  angular binning.
It also  reveals significant  alignment between red  luminous galaxies
and  large-scale  structure  out  to  $60\hMpc$ but  no  alignment  is
detected  for  blue galaxies.   The  restriction  to central  galaxies
results  in a  slight  enhancement  of the  alignment  signal for  red
galaxies,  in   particular,  at   lower  luminosities  and   at  large
separations.   However, the differences  are small,  indicating little
reduction  of  the  signal   by  satellite  galaxies  which  may  have
orientations   altered  by   the  local   tidal   field  \citep[cf.][]
{Pereira-Kuhn-05}.

The  second part of  this study  is based  on a  Millennium Simulation
semi-analytic galaxy catalog where  we have assigned an orientation to
each red, luminous and central  galaxy based on the orientation of the
mass distribution of the  inner halo. As an alternative identification
of the orientation of the  central galaxy we also used the orientation
of the host halo itself.

The mean  projected misalignment angle between  a halo as  a whole and
its  central (galaxy)  region is  $\sim 25^\circ$.   This misalignment
decreases slightly  with increasing  luminosity of the  central galaxy
and/or  its host halo  mass.  This  behavior is  particularly notable,
since the relative  size of the central region  (used to determine the
central    orientation)    decreases     with    halo    mass    (cf.,
Eq.~\ref{eqn:cooray}).

Based on  the orientations and  the luminosities of  the semi-analytic
red galaxies we compute the alignment correlation function in the same
manner as  for the  SDSS galaxy catalog.   We find that  the alignment
signal  based on the  halo orientations  overestimates that  found for
SDSS.   With the  central orientation,  however, quite  good agreement
between  the  modeled galaxy  sample  and  the  observations has  been
achieved.   In other words,  a misalignment  between halo  and central
orientations  is crucial  to matching  the observed  amplitude  of the
alignment  signal. Our  approach  to  compute the  major  axis of  the
central   elliptical  galaxy   based  on   the  central   dark  matter
distribution is  physically motivated and results in  the right amount
of  misalignment.  However,  any other  process which  leads  to equal
misalignment will be in agreement with our measurements as well.

The  large   data  volume   of  the  MS   allows  us  to   generate  a
two-dimensional map of  the alignment correlation function $w_p(x,y)$,
i.e.  we can adopt a very fine two dimensional binning and still get a
reasonable number of pair-counts for  each bin. The maps based on halo
orientation show a  substantially flattened galaxy distribution around
red  main   galaxies  out  to  at  least   $60\hMpc$.   Using  central
orientations the flattening is reduced.  In particular we have shown a
map based on red main  galaxies with $r$-band magnitudes between $-22$
and $-21$  using the standard  reference sample.  There the  ratios of
the  minor to major  axes at  $\log w_p(x,y)=1.5$,  $1$ and  $0.5$ are
0.57, 0.61  and 0.42  for halo orientations  and 0.75, 0.80,  0.75 for
central orientations.  It is interesting that the flattening is almost
independent of scale all the way from $1\hMpc$ to $60\hMpc$.
\begin{acknowledgements}
  We  would like  to thank  the referee  for valuable  comments, which
  helped to improve  the paper.  AF and CL are  supported by the Joint
  Postdoctoral  Program  in  Astrophysical  Cosmology  of  Max  Planck
  Institute  for Astrophysics  and Shanghai  Astronomical Observatory.
  AF acknowledges support from the European Science Foundation program
  `Computational  Astrophysics  and Cosmology'  and  the Jodrell  Bank
  Visitor  Grant.   YPJ  is  supported by  NSFC  (10533030,  10821302,
  10878001),  by   the  Knowledge  Innovation  Program   of  CAS  (No.
  KJCX2-YW-T05),   and   by   973   Program   (No.2007CB815402).    SM
  acknowledges the Humboldt Foundation for travel support.
  
  Funding for the SDSS and SDSS-II  has been provided by the Alfred P.
  Sloan  Foundation,  the  Participating  Institutions,  the  National
  Science  Foundation, the  U.S.  Department  of Energy,  the National
  Aeronautics and  Space Administration, the  Japanese Monbukagakusho,
  the Max Planck Society, and the Higher Education Funding Council for
  England.  The SDSS Web Site is http://www.sdss.org/.

  The SDSS is managed by the Astrophysical Research Consortium for the
  Participating Institutions.  The Participating Institutions  are the
  American Museum of Natural History, Astrophysical Institute Potsdam,
  University of  Basel, University of Cambridge,  Case Western Reserve
  University, University of  Chicago, Drexel University, Fermilab, the
  Institute for  Advanced Study, the Japan  Participation Group, Johns
  Hopkins  University, the Joint  Institute for  Nuclear Astrophysics,
  the  Kavli Institute  for Particle  Astrophysics and  Cosmology, the
  Korean Scientist  Group, the  Chinese Academy of  Sciences (LAMOST),
  Los  Alamos   National  Laboratory,  the   Max-Planck-Institute  for
  Astronomy (MPIA),  the Max-Planck-Institute for  Astrophysics (MPA),
  New Mexico  State University,  Ohio State University,  University of
  Pittsburgh,  University  of  Portsmouth, Princeton  University,  the
  United States Naval Observatory, and the University of Washington.
\end{acknowledgements}

\begin{thebibliography}{66}
\expandafter\ifx\csname natexlab\endcsname\relax\def\natexlab#1{#1}\fi

\bibitem[{{Aarseth} {et~al.}(1979){Aarseth}, {Turner}, \&
  {Gott}}]{Aarseth-Turner-Gott-79}
{Aarseth}, S.~J., {Turner}, E.~L., \& {Gott}, III, J.~R. 1979, \apj, 228, 664

\bibitem[{{Abazajian} {et~al.}(2004){Abazajian}, {Adelman-McCarthy},
  {Ag{\"u}eros}, {Allam}, {Anderson}, {Anderson}, {Annis}, {Bahcall}, \& {et
  al.,}}]{Abazajian-04}
{Abazajian}, K., {Adelman-McCarthy}, J.~K., {Ag{\"u}eros}, M.~A., {et~al.}
  2004, \aj, 128, 502

\bibitem[{{Abazajian} {et~al.}(2005){Abazajian}, {Adelman-McCarthy},
  {Ag{\"u}eros}, {Allam}, {Anderson}, {Anderson}, {Annis}, {Bahcall}, \& {et
  al.,}}]{Abazajian-05}
{Abazajian}, K., {Adelman-McCarthy}, J.~K., {Ag{\"u}eros}, M.~A., {et~al.}
  2005, \aj, 129, 1755

\bibitem[{{Abazajian} {et~al.}(2003){Abazajian}, {Adelman-McCarthy},
  {Ag{\"u}eros}, {Allam}, {Anderson}, {Annis}, {Bahcall}, {Baldry}, \& {et
  al.,}}]{Abazajian-03}
{Abazajian}, K., {Adelman-McCarthy}, J.~K., {Ag{\"u}eros}, M.~A., {et~al.}
  2003, \aj, 126, 2081

\bibitem[{{Adelman-McCarthy} {et~al.}(2008){Adelman-McCarthy}, {Ag{\"u}eros},
  {Allam}, {Allende Prieto}, {Anderson}, {Anderson}, {Annis}, {Bahcall}, \& {et
  al.,}}]{Adelman-McCarthy-08}
{Adelman-McCarthy}, J.~K., {Ag{\"u}eros}, M.~A., {Allam}, S.~S., {et~al.} 2008,
  \apjs, 175, 297

\bibitem[{{Adelman-McCarthy} {et~al.}(2007){Adelman-McCarthy}, {Ag{\"u}eros},
  {Allam}, {Anderson}, {Anderson}, {Annis}, {Bahcall}, {Bailer-Jones}, \& {et
  al.,}}]{Adelman-McCarthy-07}
{Adelman-McCarthy}, J.~K., {Ag{\"u}eros}, M.~A., {Allam}, S.~S., {et~al.} 2007,
  \apjs, 172, 634

\bibitem[{{Adelman-McCarthy} {et~al.}(2006){Adelman-McCarthy}, {Ag{\"u}eros},
  {Allam}, {Anderson}, {Anderson}, {Annis}, {Bahcall}, {Baldry}, \& {et
  al.,}}]{Adelman-McCarthy-06}
{Adelman-McCarthy}, J.~K., {Ag{\"u}eros}, M.~A., {Allam}, S.~S., {et~al.} 2006,
  \apjs, 162, 38

\bibitem[{{Agustsson} \& {Brainerd}(2006)}]{Agustsson-Brainerd-06b}
{Agustsson}, I. \& {Brainerd}, T.~G. 2006, \apj, 650, 550

\bibitem[{{Altay} {et~al.}(2006){Altay}, {Colberg}, \&
  {Croft}}]{Altay-Colberg-Croft-06}
{Altay}, G., {Colberg}, J.~M., \& {Croft}, R.~A.~C. 2006, \mnras, 370, 1422

\bibitem[{{Azzaro} {et~al.}(2007){Azzaro}, {Patiri}, {Prada}, \&
  {Zentner}}]{Azzaro-07}
{Azzaro}, M., {Patiri}, S.~G., {Prada}, F., \& {Zentner}, A.~R. 2007, \mnras,
  376, L43

\bibitem[{{Bardeen} {et~al.}(1986){Bardeen}, {Bond}, {Kaiser}, \&
  {Szalay}}]{Bardeen-86}
{Bardeen}, J.~M., {Bond}, J.~R., {Kaiser}, N., \& {Szalay}, A.~S. 1986, \apj,
  304, 15

\bibitem[{{Binggeli}(1982)}]{Binggeli-82}
{Binggeli}, B. 1982, \aap, 107, 338

\bibitem[{{Blanton} {et~al.}(2003{\natexlab{a}}){Blanton}, {Brinkmann},
  {Csabai}, {Doi}, {Eisenstein}, {Fukugita}, {Gunn}, {Hogg}, \& {et
  al.,}}]{Blanton-03b}
{Blanton}, M.~R., {Brinkmann}, J., {Csabai}, I., {et~al.} 2003{\natexlab{a}},
  \aj, 125, 2348

\bibitem[{{Blanton} {et~al.}(2003{\natexlab{b}}){Blanton}, {Lin}, {Lupton},
  {Maley}, {Young}, {Zehavi}, \& {Loveday}}]{Blanton-03a}
{Blanton}, M.~R., {Lin}, H., {Lupton}, R.~H., {et~al.} 2003{\natexlab{b}}, \aj,
  125, 2276

\bibitem[{{Blanton} {et~al.}(2005){Blanton}, {Schlegel}, {Strauss},
  {Brinkmann}, {Finkbeiner}, {Fukugita}, {Gunn}, {Hogg}, \& {et
  al.,}}]{Blanton-05}
{Blanton}, M.~R., {Schlegel}, D.~J., {Strauss}, M.~A., {et~al.} 2005, \aj, 129,
  2562

\bibitem[{{Bond} {et~al.}(1996){Bond}, {Kofman}, \&
  {Pogosyan}}]{Bond-Kofman-Pogosyan-96}
{Bond}, J.~R., {Kofman}, L., \& {Pogosyan}, D. 1996, \nat, 380, 603

\bibitem[{{Brainerd}(2005)}]{Brainerd-05}
{Brainerd}, T.~G. 2005, \apjl, 628, L101

\bibitem[{{Carter} \& {Metcalfe}(1980)}]{Carter-Metcalfe-80}
{Carter}, D. \& {Metcalfe}, N. 1980, \mnras, 191, 325

\bibitem[{{Chambers} {et~al.}(2000){Chambers}, {Melott}, \&
  {Miller}}]{Chambers-Melott-Miller-00}
{Chambers}, S.~W., {Melott}, A.~L., \& {Miller}, C.~J. 2000, \apj, 544, 104

\bibitem[{{Colberg} {et~al.}(1999){Colberg}, {White}, {Jenkins}, \&
  {Pearce}}]{Colberg-99}
{Colberg}, J.~M., {White}, S.~D.~M., {Jenkins}, A., \& {Pearce}, F.~R. 1999,
  \mnras, 308, 593

\bibitem[{{Colless} {et~al.}(2001){Colless}, {Dalton}, {Maddox}, {Sutherland},
  {Norberg}, {Cole}, {Bland-Hawthorn}, {Bridges}, \& {et al.,}}]{Colless-01}
{Colless}, M., {Dalton}, G., {Maddox}, S., {et~al.} 2001, \mnras, 328, 1039

\bibitem[{{Cooray} \& {Milosavljevi{\'c}}(2005)}]{Cooray-Milosavljevic-05}
{Cooray}, A. \& {Milosavljevi{\'c}}, M. 2005, \apjl, 627, L85

\bibitem[{{Croft} \& {Metzler}(2000)}]{Croft-Metzler-00}
{Croft}, R.~A.~C. \& {Metzler}, C.~A. 2000, \apj, 545, 561

\bibitem[{{Croton} {et~al.}(2006){Croton}, {Springel}, {White}, {De Lucia},
  {Frenk}, {Gao}, {Jenkins}, {Kauffmann}, \& {et al.,}}]{Croton-06}
{Croton}, D.~J., {Springel}, V., {White}, S.~D.~M., {et~al.} 2006, \mnras, 365,
  11

\bibitem[{{Davis} {et~al.}(1985){Davis}, {Efstathiou}, {Frenk}, \&
  {White}}]{Davis-85}
{Davis}, M., {Efstathiou}, G., {Frenk}, C.~S., \& {White}, S.~D.~M. 1985, \apj,
  292, 371

\bibitem[{{De Lucia} \& {Blaizot}(2007)}]{DeLucia-Blaizot-07}
{De Lucia}, G. \& {Blaizot}, J. 2007, \mnras, 375, 2

\bibitem[{{Dekel}(1985)}]{Dekel-85}
{Dekel}, A. 1985, \apj, 298, 461

\bibitem[{{Donoso} {et~al.}(2006){Donoso}, {O'Mill}, \&
  {Lambas}}]{Donoso-O'Mill-Lambas-06}
{Donoso}, E., {O'Mill}, A., \& {Lambas}, D.~G. 2006, \mnras, 369, 479

\bibitem[{{Eke} {et~al.}(2001){Eke}, {Navarro}, \&
  {Steinmetz}}]{Eke-Navarro-Steinmetz-01}
{Eke}, V.~R., {Navarro}, J.~F., \& {Steinmetz}, M. 2001, \apj, 554, 114

\bibitem[{{Faltenbacher} {et~al.}(2008){Faltenbacher}, {Jing}, {Li}, {Mao},
  {Mo}, {Pasquali}, \& {van den Bosch}}]{Faltenbacher-08}
{Faltenbacher}, A., {Jing}, Y.~P., {Li}, C., {et~al.} 2008, \apj, 675, 146

\bibitem[{{Faltenbacher} {et~al.}(2007){Faltenbacher}, {Li}, {Mao}, {van den
  Bosch}, {Yang}, {Jing}, {Pasquali}, \& {Mo}}]{Faltenbacher-07}
{Faltenbacher}, A., {Li}, C., {Mao}, S., {et~al.} 2007, \apjl, 662, L71

\bibitem[{{Fukugita} {et~al.}(1996){Fukugita}, {Ichikawa}, {Gunn}, {Doi},
  {Shimasaku}, \& {Schneider}}]{Fukugita-96}
{Fukugita}, M., {Ichikawa}, T., {Gunn}, J.~E., {et~al.} 1996, \aj, 111, 1748

\bibitem[{{Hashimoto} {et~al.}(2008){Hashimoto}, {Henry}, \&
  {Boehringer}}]{Hashimoto-Henry-Boehringer-08}
{Hashimoto}, Y., {Henry}, J.~P., \& {Boehringer}, H. 2008, \mnras, 390, 1562

\bibitem[{{Hashimoto} {et~al.}(2007){Hashimoto}, {Henry}, \&
  {B{\"o}hringer}}]{Hashimoto-Henry-Bohringer-07}
{Hashimoto}, Y., {Henry}, J.~P., \& {B{\"o}hringer}, H. 2007, \mnras, 380, 835

\bibitem[{{Heavens} {et~al.}(2000){Heavens}, {Refregier}, \&
  {Heymans}}]{Heavens-Refregier-Heymans-00}
{Heavens}, A., {Refregier}, A., \& {Heymans}, C. 2000, \mnras, 319, 649

\bibitem[{{Hirata} {et~al.}(2007){Hirata}, {Mandelbaum}, {Ishak}, {Seljak},
  {Nichol}, {Pimbblet}, {Ross}, \& {Wake}}]{Hirata-07}
{Hirata}, C.~M., {Mandelbaum}, R., {Ishak}, M., {et~al.} 2007, \mnras, 381,
  1197

\bibitem[{{Ivezi{\'c}} {et~al.}(2004){Ivezi{\'c}}, {Lupton}, {Schlegel},
  {Boroski}, {Adelman-McCarthy}, {Yanny}, {Kent}, {Stoughton}, \& {et
  al.,}}]{Ivezic-04}
{Ivezi{\'c}}, {\v Z}., {Lupton}, R.~H., {Schlegel}, D., {et~al.} 2004,
  Astronomische Nachrichten, 325, 583

\bibitem[{{Jing}(2002)}]{Jing-02}
{Jing}, Y.~P. 2002, \mnras, 335, L89

\bibitem[{{Jing} {et~al.}(1998){Jing}, {Mo}, \& {Boerner}}]{Jing-Mo-Boerner-98}
{Jing}, Y.~P., {Mo}, H.~J., \& {Boerner}, G. 1998, \apj, 494, 1

\bibitem[{{Kang} {et~al.}(2007){Kang}, {van den Bosch}, {Yang}, {Mao}, {Mo},
  {Li}, \& {Jing}}]{Kang-07}
{Kang}, X., {van den Bosch}, F.~C., {Yang}, X., {et~al.} 2007, \mnras, 378,
  1531

\bibitem[{{Knebe} {et~al.}(2008){Knebe}, {Yahagi}, {Kase}, {Lewis}, \&
  {Gibson}}]{Knebe-08a}
{Knebe}, A., {Yahagi}, H., {Kase}, H., {Lewis}, G., \& {Gibson}, B.~K. 2008,
  \mnras, 388, L34

\bibitem[{{Li} {et~al.}(2006){Li}, {Kauffmann}, {Jing}, {White}, {B{\"o}rner},
  \& {Cheng}}]{Li-06b}
{Li}, C., {Kauffmann}, G., {Jing}, Y.~P., {et~al.} 2006, \mnras, 368, 21

\bibitem[{{Lupton} {et~al.}(2001){Lupton}, {Gunn}, {Ivezi{\'c}}, {Knapp}, \&
  {Kent}}]{Lupton-01}
{Lupton}, R., {Gunn}, J.~E., {Ivezi{\'c}}, Z., {Knapp}, G.~R., \& {Kent}, S.
  2001, in Astronomical Data Analysis Software and Systems X, ed. F.~R.
  {Harnden}, Jr., F.~A. {Primini}, \& H.~{Payne}, Vol. 238, 269--+

\bibitem[{{Mandelbaum} {et~al.}(2006){Mandelbaum}, {Seljak}, {Cool}, {Blanton},
  {Hirata}, \& {Brinkmann}}]{Mandelbaum-06a}
{Mandelbaum}, R., {Seljak}, U., {Cool}, R.~J., {et~al.} 2006, \mnras, 372, 758

\bibitem[{{Miralda-Escude}(1991)}]{Miralda-Escude-91}
{Miralda-Escude}, J. 1991, \apj, 380, 1

\bibitem[{{Mo} \& {White}(1996)}]{Mo-White-96}
{Mo}, H.~J. \& {White}, S.~D.~M. 1996, \mnras, 282, 347

\bibitem[{{Okumura} {et~al.}(2008){Okumura}, {Jing}, \&
  {Li}}]{Okumura-Jing-Li-08}
{Okumura}, T., {Jing}, Y.~P., \& {Li}, C. 2008, ArXiv e-prints

\bibitem[{{Paz} {et~al.}(2008){Paz}, {Stasyszyn}, \&
  {Padilla}}]{Paz-Stasyszyn-Padilla-08}
{Paz}, D., {Stasyszyn}, F., \& {Padilla}, N. 2008, ArXiv e-prints

\bibitem[{{Peebles}(1980)}]{Peebles-80}
{Peebles}, P.~J.~E. 1980, {The large-scale structure of the universe} (Research
  supported by the National Science Foundation.~Princeton, N.J., Princeton
  University Press, 1980.~435 p.)

\bibitem[{{Pereira} \& {Kuhn}(2005)}]{Pereira-Kuhn-05}
{Pereira}, M.~J. \& {Kuhn}, J.~R. 2005, \apjl, 627, L21

\bibitem[{{Plionis}(1994)}]{Plionis-94}
{Plionis}, M. 1994, \apjs, 95, 401

\bibitem[{{Plionis} {et~al.}(2003){Plionis}, {Benoist}, {Maurogordato},
  {Ferrari}, \& {Basilakos}}]{Plionis-03}
{Plionis}, M., {Benoist}, C., {Maurogordato}, S., {Ferrari}, C., \&
  {Basilakos}, S. 2003, \apj, 594, 144

\bibitem[{{Smith} {et~al.}(2002){Smith}, {Tucker}, {Kent}, {Richmond},
  {Fukugita}, {Ichikawa}, {Ichikawa}, {Jorgensen}, \& {et al.,}}]{Smith-02}
{Smith}, J.~A., {Tucker}, D.~L., {Kent}, S., {et~al.} 2002, \aj, 123, 2121

\bibitem[{{Springel}(2005)}]{Springel-05a}
{Springel}, V. 2005, \mnras, 364, 1105

\bibitem[{{Springel} {et~al.}(2005){Springel}, {White}, {Jenkins}, {Frenk},
  {Yoshida}, {Gao}, {Navarro}, {Thacker}, \& {et al.,}}]{Springel-05b}
{Springel}, V., {White}, S.~D.~M., {Jenkins}, A., {et~al.} 2005, \nat, 435, 629

\bibitem[{{Springel} {et~al.}(2001){Springel}, {White}, {Tormen}, \&
  {Kauffmann}}]{Springel-01}
{Springel}, V., {White}, S.~D.~M., {Tormen}, G., \& {Kauffmann}, G. 2001,
  \mnras, 328, 726

\bibitem[{{Stoughton} {et~al.}(2002){Stoughton}, {Lupton}, {Bernardi},
  {Blanton}, {Burles}, {Castander}, {Connolly}, {Eisenstein}, \& {et
  al.,}}]{Stoughton-02}
{Stoughton}, C., {Lupton}, R.~H., {Bernardi}, M., {et~al.} 2002, \aj, 123, 485

\bibitem[{{Struble}(1990)}]{Struble-90}
{Struble}, M.~F. 1990, \aj, 99, 743

\bibitem[{{Ulmer} {et~al.}(1989){Ulmer}, {McMillan}, \&
  {Kowalski}}]{Ulmer-McMillan-Kowalski-89}
{Ulmer}, M.~P., {McMillan}, S.~L.~W., \& {Kowalski}, M.~P. 1989, \apj, 338, 711

\bibitem[{{Wang} {et~al.}(2008){Wang}, {Yang}, {Mo}, {Li}, {van den Bosch},
  {Fan}, \& {Chen}}]{Wang-08}
{Wang}, Y., {Yang}, X., {Mo}, H.~J., {et~al.} 2008, \mnras, 385, 1511

\bibitem[{{West}(1989{\natexlab{a}})}]{West-89b}
{West}, M.~J. 1989{\natexlab{a}}, \apj, 344, 535

\bibitem[{{West}(1989{\natexlab{b}})}]{West-89a}
{West}, M.~J. 1989{\natexlab{b}}, \apj, 347, 610

\bibitem[{{Yang} {et~al.}(2005{\natexlab{a}}){Yang}, {Mo}, {Jing}, \& {van den
  Bosch}}]{Yang-05b}
{Yang}, X., {Mo}, H.~J., {Jing}, Y.~P., \& {van den Bosch}, F.~C.
  2005{\natexlab{a}}, \mnras, 358, 217

\bibitem[{{Yang} {et~al.}(2005{\natexlab{b}}){Yang}, {Mo}, {van den Bosch}, \&
  {Jing}}]{Yang-05a}
{Yang}, X., {Mo}, H.~J., {van den Bosch}, F.~C., \& {Jing}, Y.~P.
  2005{\natexlab{b}}, \mnras, 356, 1293

\bibitem[{{Yang} {et~al.}(2006){Yang}, {van den Bosch}, {Mo}, {Mao}, {Kang},
  {Weinmann}, {Guo}, \& {Jing}}]{Yang-06b}
{Yang}, X., {van den Bosch}, F.~C., {Mo}, H.~J., {et~al.} 2006, \mnras, 369,
  1293

\bibitem[{{York} {et~al.}(2000){York}, {Adelman}, {Anderson}, {Anderson},
  {Annis}, {Bahcall}, {Bakken}, {Barkhouser}, \& {et al.,}}]{York-00}
{York}, D.~G., {Adelman}, J., {Anderson}, Jr., J.~E., {et~al.} 2000, \aj, 120,
  1579

\end{thebibliography}


\label{lastpage}

\end{document}